\documentclass[%
 reprint,
superscriptaddress,
 amsmath,amssymb,
 aps,
 prl,
]{revtex4-2}

\usepackage{graphicx}% Include figure files
\usepackage{dcolumn}% Align table columns on decimal point
\usepackage{bm}% bold math
\usepackage{amsmath}
\usepackage{slashed}
\usepackage{url}
\usepackage{hyperref}% add hypertext capabilities
\usepackage{xcolor}
\usepackage{comment}
\usepackage{enumitem}
\usepackage{blkarray}
\newcommand{\cev}[1]{\reflectbox{\ensuremath{\vec{\reflectbox{\ensuremath{#1}}}}}}
\newcommand{\Tr}[0]{{\rm Tr}}

\begin{document}

\preprint{APS/123-QED}

\author{Simone Bacchio}
\affiliation{Computation-based Science and Technology Research Center,
The Cyprus Institute, 20 Kavafi Str., Nicosia 2121, Cyprus}
\author{Andreas Konstantinou}
\affiliation{Department of Physics, University of Cyprus, P.O. Box 20537, 1678 Nicosia, Cyprus}

\title{Study of the $\Lambda \rightarrow p\ell \bar{\nu}_\ell$ semileptonic decay in lattice QCD}

\begin{abstract}
We present the first lattice QCD determination of the $\Lambda \to N$ vector and axial-vector form factors, which are essential inputs for studying the semileptonic decay $\Lambda \to p \ell \bar{\nu}_\ell$. This channel provides a clean, theoretically controlled avenue for extracting the CKM matrix element $|V_{us}|$ from the baryon sector. Our analysis uses a gauge ensemble with physical light, strange, and charm quark masses and yields the most precise determination to date of the full set of transition form factors—including second-class contributions—as well as the associated couplings, radii, and the ratio of muon-to-electron decay rates, an observable sensitive to possible non-standard scalar and tensor interactions.
We compare our non-perturbative results with next-to-next-to-leading order expansions in the small parameter $\delta = (m_\Lambda - m_N)/m_\Lambda \approx 0.16$. We find that the common phenomenological approximation of neglecting the $q^2$-dependence of the form factors leads to a $\sim 4\%$ deviation in the decay rate. This underscores the critical importance of precise, fully non-perturbative form factor inputs for achieving the sub-percent precision targets of upcoming experimental programs.
\end{abstract}


\maketitle

\section{Introduction} \label{sec:intro}

The semileptonic decay $\Lambda \to p\ell \bar{\nu}_\ell$ is one of the weak decay modes of the $\Lambda$ hyperon, where the valence strange quark of the $\Lambda$ transforms into an up quark via the weak interaction  emitting a virtual $W$-boson that subsequently decays into a lepton-neutrino pair. This process provides a valuable probe of the weak interactions and hadron structure. 
From a theoretical perspective, the total decay amplitude can be factorized into two components: the hadronic matrix element, describing the $\Lambda \to p$ transition and the leptonic matrix element. The resulting decay amplitude is 
\begin{equation}\label{eq:M}
\mathcal{M} = \frac{G_F}{\sqrt{2}}\, V_{us}\, \langle p | \bar{u}\,\varGamma^\text{V-A}_\mu s | \Lambda \rangle \cdot \langle \ell \bar{\nu}_\ell | \bar{\ell}\,\varGamma^\text{V-A}_\mu \nu_\ell | 0 \rangle\,,
\end{equation} 
where $\varGamma^\text{V-A}_\mu = \gamma_\mu (1 - \gamma_5)$ is the left-handed operator representing in the Standard Model the weak coupling between left-handed fermions. Two components in this expression are well known: the Fermi constant, $G_F$, which is determined from muon decay~\cite{MuLan:2007qkz}, and the leptonic matrix elements, $\langle \ell \bar{\nu}_\ell | \bar{\ell}\,\varGamma^\text{V-A}_\mu \nu_\ell | 0 \rangle$, which are computed perturbatively. The remaining two components are less known and constitute the central focus of this work.

\textbullet{} The hadronic matrix element  
   $
   \langle p | \bar{u}\,\varGamma^\text{V-A}_\mu s | \Lambda \rangle
   $  
   encodes the non-perturbative QCD dynamics governing the transition of a $\Lambda$ hyperon into a proton. It is conventionally expressed in terms of six form factors, which encapsulate the effects of strong interactions. While previous phenomenological studies have extracted these form factors using a combination of experimental data, effective models, and symmetry arguments, first-principles calculations from lattice QCD have not been performed prior to this work for this particular decay.  

\textbullet{} The Cabibbo–Kobayashi–Maskawa (CKM) matrix element $V_{us}$—related to the Cabibbo angle by $V_{us}{\equiv} \sin(\theta_{\rm Cabibbo})$—governs the strength of weak transitions between strange and up quarks through its magnitude $|V_{us}|$. Currently, this value exhibits a tension among different approaches used for its determination, such as extractions from kaon and tau decays, as well as from the CKM first-row unitarity relation~\cite{ParticleDataGroup:2024cfk}. Given this discrepancy, independent determinations of $|V_{us}|$ are of critical importance.

Semi-leptonic hyperon decays, such as $\Lambda \to p\ell \bar{\nu}_\ell$, offer an alternative and complementary approach to extracting $|V_{us}|$, as first studied by Cabibbo {\it et al.} in Ref.~\cite{Cabibbo:2003ea,Cabibbo:2003cu}. However, due to the absence of updated experimental measurements and non-perturbative theoretical calculations, this two-decade-old result remains the state-of-the-art determination from hyperons~\cite{ParticleDataGroup:2024cfk}. 
By providing a first-principles lattice QCD determination of the $\Lambda \to N$ form factors, this work establishes the basis for a more precise and accurate extraction of $|V_{us}|$ from hyperon decays. The calculation is performed on a single gauge ensemble generated at physical light, strange, and charm quark masses by the Extended Twisted Mass Collaboration (ETMC).  This study clearly demonstrates that lattice QCD can compute the relevant matrix elements with sufficient precision to enable a fully controlled, non-perturbative determination of $|V_{us}|$ from hyperon decays. 

 This work  aligns with current experimental efforts to measure the semileptonic decay $\Lambda \to p\ell \bar{\nu}_\ell$ at BESIII and LHCb. While previous hyperon decay studies~\cite{ParticleDataGroup:2024cfk} focused solely on the decay process, excluding the production mechanism, BESIII’s ability to reconstruct the full baryon–antibaryon ($B\bar{B}$) pair enables a novel approach that incorporates the production process~\cite{Batozskaya:2023rek}. This allows simultaneous modeling of the joint angular distributions of production and decay, providing additional observables in various kinematic regions to compare with theoretical predictions. Furthermore, a comparison between first-principles Standard Model calculations and experimental measurements provides constraints on non-standard interactions~\cite{Cirigliano:2009wk,Cirigliano:2012ab,Chang:2014iba}. Deviations in the axial coupling may signal non-zero right-handed currents, while discrepancies in the ratio of the muon- to electron-channel decay rate, $\Gamma(\Lambda \to p\mu \bar{\nu}_\mu)/\Gamma(\Lambda \to pe \bar{\nu}_e)$, may indicate scalar and/or tensor contributions~\cite{Chang:2014iba}.

\section{Theoretical Formalism}\label{sec:FFs}

The rate $\Gamma$ of the semileptonic decay $\Lambda \to p \, \ell \, \bar{\nu}_\ell$ can be computed as an integral over the final-state phase space,
\begin{multline}
    \Gamma = \frac{1}{2m_\Lambda} \int \frac{d^3\vec{p}_p}{(2\pi)^3 2E_p} \frac{d^3\vec{p}_\ell}{(2\pi)^3 2E_\ell} \frac{d^3\vec{p}_\nu}{(2\pi)^3 2E_\nu}\times\\ (2\pi)^4 \delta^{(4)}(p_\Lambda - p_p - p_\ell - p_\nu) \, \left| \mathcal{M} \right|^2,
\end{multline}
where $\mathcal{M}$ is the invariant matrix element defined in Eq.~\eqref{eq:M}, $p$ is the 4-momentum and $E=\sqrt{m^2 + |\vec{p}\,|^2}$ are the energies of the initial and final particles of mass $m$. Integrating out the angular variables leads to a dependence only on the squared momentum transfer $q^2 = (p_\Lambda - p_p)^2 = (p_\ell + p_\nu)^2$, which corresponds to the invariant mass squared of the lepton-neutrino system. The decay rate then reduces to
\begin{equation}\label{eq:decay_rate}
\Gamma = \frac{G_F^2 |V_{us}|^2}{192\pi^3m_\Lambda^3} \int_{m_\ell^2}^{q^2_{\max}} dq^2 \sqrt{\lambda(m_\Lambda^2, m_p^2, q^2)}\, L^{\mu\nu} H_{\mu\nu} ,
\end{equation}
where $q^2_{\max} = (m_{\Lambda} - m_p)^2$ is the maximum allowed momentum transfer, $\lambda(a,b,c) = a^2 + b^2 + c^2 - 2(ab + bc + ca)$ is the K\"allén function and $L^{\mu\nu}$ and $H_{\mu\nu}$ are the leptonic and hadronic tensors, respectively. The latter, arising from the angular integration of $\left| \mathcal{M} \right|^2$, are given by
\begin{align}
    L^{\mu\nu} &\equiv \left( 1 - \frac{m_\ell^2}{q^2} \right)^2\left[\frac{q^\mu q^\nu}{q^2}\left(1+\frac{m_\ell^2}{2q^2}\right)-g^{\mu\nu}\right]\,,\\
    H_{\mu\nu}&\equiv\sum_{\rm s_p,s_\Lambda} M_\mu^\text{V-A}\left(M_\nu^\text{V-A}\right)^*\,,
\end{align}
where $g_{\mu\nu}=\texttt{diag(+,-,-,-)}$ is the Minkowski metric, $s_p,s_\Lambda$ are the spins of the baryons and the hadronic matrix element $M_\mu^\text{V-A}$ is defined as
\begin{equation}\label{eq:ME}
    M_\mu^\text{V-A} \equiv \langle p (p_p,s_p)| \bar{u}\,\varGamma^\text{V-A}_\mu s | \Lambda(p_\Lambda,s_\Lambda) \rangle\,.
\end{equation}

    \noindent
{\bf From matrix elements to form factors.} 
The hadronic matrix elements are expressed in terms of Lorentz-invariant form factors, which provide a general decomposition consistent with the symmetries of QCD and depend only on the momentum transfer $q^2$. In this work, we use both the Weinberg~\cite{Gutsche:2014zna} and helicity~\cite{Feldmann:2011xf} formalisms: the former is standard in phenomenological analyses, while the latter is better suited for computing the decay rate, as it maps directly onto angular distributions and polarization observables. The decomposition of the matrix elements in the Weinberg basis and Minkowski space is given by
\begin{align}\label{eq:FF}
        &~M_\mu^\text{V-A} =  
\bar{u}_p \Bigg[ \gamma_\mu\left(F_1(q^2)-G_1(q^2)\gamma_5\right)  +\\ &-i \sigma_{\mu\nu} q^\nu\frac{F_2(q^2)-G_2(q^2)\gamma_5}{m_{\Lambda}}  + q_\mu \frac{F_3(q^2)-G_3(q^2)\gamma_5}{m_{\Lambda}}  \Bigg] u_{\Lambda}\nonumber
\end{align}
where $\sigma^{\mu \nu}=\frac{i}{2}[\gamma^\mu,\gamma^\nu]$ is the antisymmetric tensor Dirac matrix and $u_{\Lambda},u_{p}$ are spinors defined as
\begin{equation}
    u_B\equiv\frac{\slashed{p}{}_B+m_B}{\sqrt{2E_B(E_B+m_B)}} u^{s},
\end{equation}
for $B\in\{p,\Lambda\}$ with $u^s$ being an up- or down-spin state. 

The form factors  $F_1(q^2),\, F_2(q^2)$ and $F_3(q^2)$ originate from the vector current $V_\mu = \bar{u}\gamma_\mu s$, while $G_1(q^2),\, G_2(q^2)$ and $G_3(q^3)$ arise from the axial-vector current $A_\mu = \bar{u}\gamma_\mu \gamma_5s$.  $F_1(q^2)$ and $F_2(q^2)$ are known as the Dirac and Pauli form factors, and $G_1(q^2)$ and $G_3(q^2)$ as the axial and induced pseudoscalar form factors—structures also present in diagonal matrix elements.
In contrast, $F_3(q^2)$ and $G_2(q^2)$ are second-class form factors~\cite{Weinberg:1958ut}: they are proportional to the mass splitting and vanish in the exact $SU(3)$ flavor symmetry limit, where the quark masses are degenerate and $m_p = m_\Lambda$. Additionally, both $F_3(q^2)$ and $G_3(q^2)$ contribute to the decay rate only through terms suppressed by $m_\ell^2$. As a result, $F_3(q^2)$ is negligible in all channels at the targeted precision, while $G_3(q^2)$ becomes relevant only in the muonic mode, since $m_\mu$ is similar in size to the mass splitting.

Further details on the theoretical formalism are provided in the appendix, including the definitions of the helicity form factors, the relations between Weinberg and helicity, the expression for the decay rate once the leptonic and hadronic tensors are contracted, as well as the perturbative approximation of the decay rate at leading orders of the mass splitting.

\section{Lattice QCD Methodology} \label{sec:methodology}

Lattice Quantum Chromodynamics (QCD) provides a non-perturbative framework for studying the strong interaction by discretizing the QCD Lagrangian on a four-dimensional Euclidean spacetime lattice. This enables \textit{ab initio} numerical simulations that incorporate the full dynamics of QCD with controlled and systematically improvable uncertainties. In the past decade, lattice QCD has reached high precision across a wide range of observables and extended into regimes previously out of reach. Of particular relevance to this study are precise lattice results on nucleon structure, including isovector charges~\cite{Walker-Loud:2019cif,Bali:2023sdi,Djukanovic:2024krw,Alexandrou:2024ozj,FlavourLatticeAveragingGroupFLAG:2024oxs,Jang:2023zts}, axial form factors~\cite{Jang:2023zts,Djukanovic:2022wru,Alexandrou:2023qbg}, and electromagnetic form factors~\cite{Djukanovic:2023beb,Tsuji:2023llh,Alexandrou:2025vto}. These advances reflect significant progress in controlling key systematics, including excited-state contamination, discretization effects, and simulations at the physical quark masses.

The lattice approach is especially well suited for studying semileptonic baryon decays via form factor determinations, as demonstrated in several prior calculations: $\Lambda_b \to N, \Lambda_c$~\cite{Detmold:2015aaa}, $\Lambda_c \to \Lambda$~\cite{Meinel:2016dqj}, $\Lambda_c \to N$~\cite{Meinel:2017ggx}, $\Lambda_c \to \Lambda^{(1520)}$~\cite{Meinel:2021mdj}, $\Lambda_b \to \Lambda_c^{(2595,2625)}$~\cite{Meinel:2021rbm}, and $\Xi_c \to \Xi$~\cite{Farrell:2025gis}. In contrast, lattice QCD results for hyperon transitions remain sparse, limited to the vector form factor $F_1(q^2)$ in $\Sigma \to N$ and $\Xi \to \Sigma$ transition~\cite{Shanahan:2015dka,Sasaki:2017jue}, or to early studies of form factors performed at heavier-than-physical pion masses for the $\Xi \to \Sigma$ transition~\cite{Sasaki:2008ha}. The present work represents a significant step forward, extending the scope and phenomenological impact of lattice studies of hyperon semileptonic decays.

\smallskip
\noindent
{\bf Simulation setup.}
We analyze a gauge ensemble generated with light, strange, and charm quark masses set to their physical values, referred to as the \textit{physical point}. Working at the physical point eliminates the need for a chiral extrapolation in the pion mass, a significant advantage. As the aim of this work is to establish the theoretical framework and demonstrate that lattice QCD can achieve the precision required for an impactful determination of the form factors involved in this decay and the precise extraction of $|V_{us}|$, we perform the analysis using a single gauge ensemble with lattice spacing $a\simeq0.08$\,fm and physical volume $L\simeq5$\,fm. Lattice systematic effects related to discretization and finite volume are not addressed in the present work. Previous studies of nucleon structure on the same ensemble~\cite{Alexandrou:2024ozj,Alexandrou:2023qbg,Alexandrou:2025zuj} suggest that such effects are small, in part thanks to the use of an $\mathcal{O}(a)$-improved action. Nevertheless, we plan to quantify these effects in future work to provide a complete error budget. The simulations are performed in the SU(2)-isosymmetric limit, where up and down quarks are degenerate, rendering the proton and neutron degenerate as well, collectively referred to as nucleons, $N$. Leading-order isospin-breaking and radiative corrections are expected to be small, contributing at the (sub-)percent level, and are neglected in current state-of-the-art lattice calculations of baryon matrix elements.

\smallskip
\noindent
{\bf Interpolating fields and correlation functions.}
To isolate the nucleon and $\Lambda$-hyperon ground-states, we employ the standard interpolating fields 
\begin{align}
\chi_N(x) &= \epsilon^{abc} \left(u^a(x)^T C\gamma_5 d^b(x)\right) u^c(x)\,,\\[0.2cm]
\chi_\Lambda(x)&=\frac{1}{\sqrt{6}}\epsilon^{abc} \Big[2\big(u^T_a(x) C\gamma_5d^b(x)\big)s^c(x)+\\&\big(u^a(x)^T C\gamma_5s^b(x)\big)d^c(x)-\big(d^a(x)^T C\gamma_5s^b(x)\big)u^c(x)\Big]\,,\nonumber
\end{align}
where $u$, $d$, and $s$ denote the up-, down-, and strange-quark spinors at the spacetime point $x\equiv(\vec{x}, t)$, $C$ is the charge conjugation matrix, and $a$, $b$, $c$ are implicitly summed color indices.
We construct two-point correlation functions as
\begin{equation}\label{eq:2pt}
C_B(\vec{p}_B,t_{\rm s})=
\sum_{\vec{x}_s} \Tr\left[P_0\langle \chi_B(x_s) \bar{\chi}_B(0)\rangle e^{-i\vec{x}_s\cdot \vec{p}_B}\right],
\end{equation}
for $B\in\{N,\Lambda\}$, where the trace is taken over the implicit Dirac indices of the interpolating fields, and the $\Lambda\to N$ transition three-point correlation function as
\begin{multline}\label{eq:3pt}
C^\text{V-A}_{\mu\nu}(\vec{p}_N,\vec{p}_\Lambda,t_{\rm s},t_{\rm ins})= \sum_{\vec{x}_{\rm ins},\vec{x}_s} e^{i\vec{x}_{\rm ins}\cdot (\vec{p}_\Lambda-\vec{p}_N)-i\vec{x}_{s}\cdot \vec{p}_\Lambda}\times\\\Tr\left[P_\nu\langle \chi_\Lambda(x_s) \bar{u}(x_{\rm ins})\,\varGamma^\text{V-A}_\mu s(x_{\rm ins})\bar{\chi}_N(0)\rangle \right].
\end{multline}
Here, $x_s$ and $x_{\text{ins}}$ denote the spacetime coordinates of the $\Lambda$ annihilation (sink) and current insertion, respectively, relatively to the spacetime of the creation of the nucleon  (source) taken, without loss of generality, at zero. Two-point functions are Fourier-transformed at the sink to access various momenta $\vec{p}_B$, while three-point functions are constructed using the fixed-sink method at fixed $t_{\rm s}$ and sink momentum $\vec{p}_\Lambda {=} \vec{0}$, and Fourier-transformed with respect to the  momentum transfer, $\vec{q}$. $P_\nu$ is a projector, which acts on spin indices. For unpolarized matrix elements with positive parity we use $P_0 = \frac{1}{2}(1 + \gamma_0)$, while for polarized matrix elements we use $P_k = iP_0 \gamma_5\gamma_k$.

\smallskip
\noindent
{\bf Extraction of ground-state matrix elements.}
Matrix elements are extracted from appropriately defined ratios of two- and three-point functions that cancel both the overlap factors between interpolating fields and hadron states, and  exponential time dependencies. We use
\begin{equation}
\label{eq:ratio}
R^\text{V-A}_{\mu\nu}(\vec{p}_N,\vec{p}_\Lambda,t_{\rm s},t_{\rm ins})\equiv \frac{C^\text{V-A}_{\mu\nu}(\vec{p}_N,\vec{p}_\Lambda,t_{\rm s},t_{\rm ins})}{\sqrt{C_\Lambda(\vec{p}_\Lambda,t_{\rm s})
C_N(\vec{p}_N,t_{\rm s})
}}\,.
\end{equation}
In the limit of large time separations, $t_{\rm s} \gg a$, and for operator insertion at $t_{\rm ins} = t_{\rm s}/2$, the ratio becomes time-independent,
\begin{align}\label{eq:LME}
R^\text{V-A}_{\mu\nu}(\vec{p}_N,\vec{p}_\Lambda,t_{\rm s},t_{\rm s}/2) \xrightarrow[t_{s} \gg a]{}\Pi^\text{V-A}_{\mu\nu}(\vec{p}_N,\vec{p}_\Lambda)\,,
\end{align}
obtaining matrix elements like those defined in Eq.~\eqref{eq:FF}, but for a given polarization $P_\nu$ and momentum choice.
Form factors are extracted from linear combinations of $\Pi^\text{V-A}_{\mu\nu}(\vec{p}_N,\vec{p}_\Lambda)$. A critical aspect of reliably determining ground-state matrix elements is controlling excited-state contamination. Achieving ground-state dominance is hindered by the rapid growth of statistical noise with increasing source-sink separation $t_{\rm s}$. To mitigate this, we employ multiple strategies, including the summation method and two-state fits to model the time-dependence of the ration in Eq.~\eqref{eq:ratio} and determine its asymptotic behavior. The full extraction procedure, along with a detailed account of the systematics, is provided in the supplementary material.

\section{Results}\label{results}

\begin{figure}
\centering
{\includegraphics[width = \linewidth]{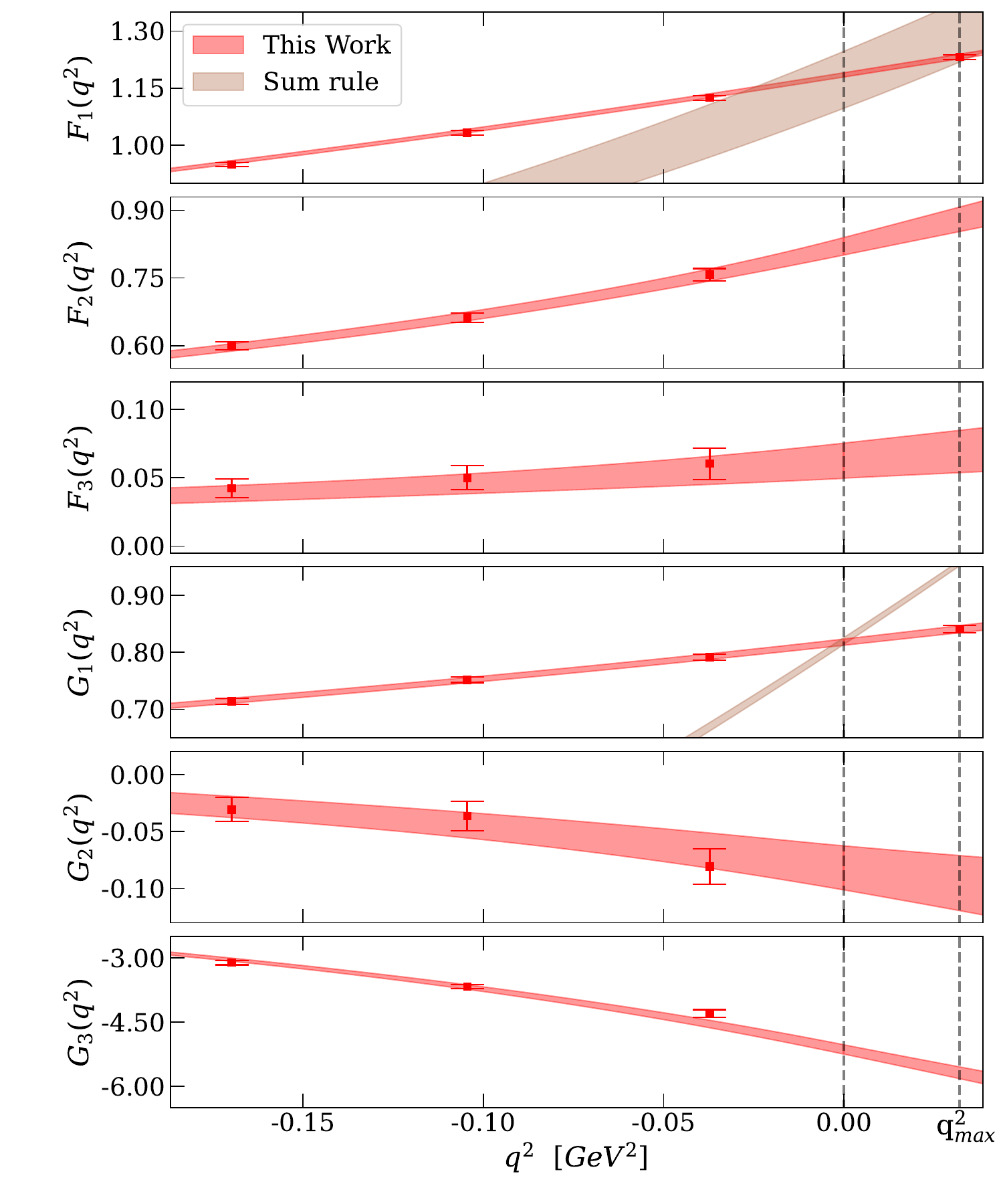}}
\caption{Momentum transfer dependence of the Weinberg form factors. Red points denote selected lattice data. The red band represents the $z$-expansion fit, while the brown band shows results from QCD sum rules~\cite{Zhang:2024ick}. Vertical dashed lines indicate the kinematic region relevant for the semileptonic decay rate, $q^2 \in [0, q^2_{\max}]$.
}
\label{fig:WFF}
\end{figure}

\begin{figure}
\centering
{\includegraphics[width = \linewidth]{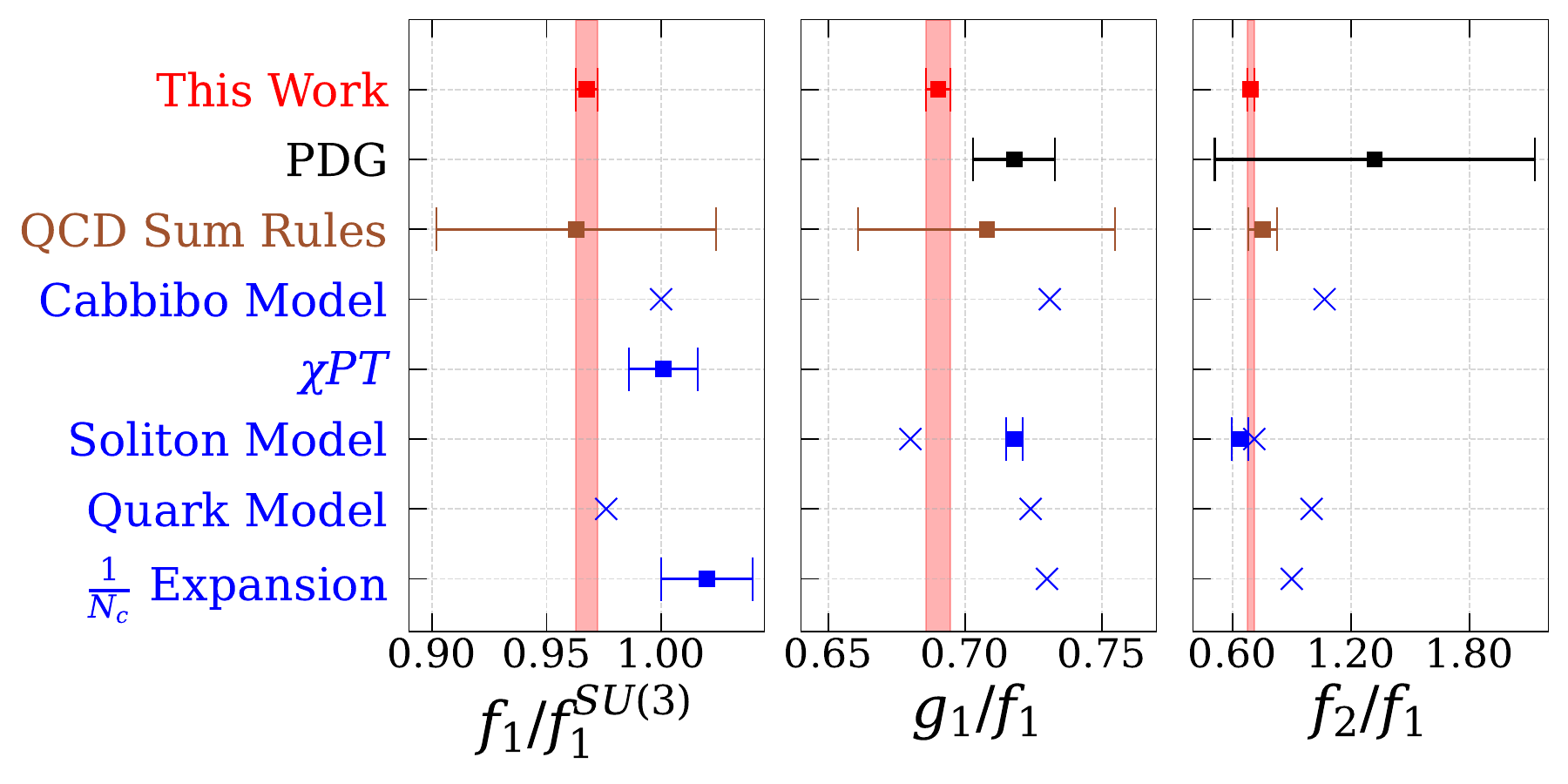}} 
\caption{Comparison of  ratios of coupling constants with experimental  and phenomenological ones. Points marked with a cross do not have uncertainties. The experimental values for $g_1/f_1$ and $f_2/f_1$ are taken from the Particle Data Group (PDG)~\cite{ParticleDataGroup:2024cfk} and are shown in black. Results from phenomenology  include QCD sum rules~\cite{Zhang:2024ick},  Cabibbo's model~\cite{Cabibbo:2003cu}, chiral perturbation theory ($\chi$PT)~\cite{Geng:2009ik}, the soliton model—where points with errors are from Ref.~\cite{Yang:2015era} and those without from Ref.~\cite{Ledwig:2008ku}—the quark model—where $f_1$ is taken from Ref.~\cite{Schlumpf:1994fb}, and $g_1/f_1$ and $f_2/f_1$ from Ref.~\cite{Faessler:2008ix}—and the $1/N_c$ expansion~\cite{Flores-Mendieta:1998tfv}.}
\label{fig:comp}
\end{figure}

\noindent
{\bf Weinberg form factors and couplings.}
In Fig.~\ref{fig:WFF}, we show our final results for the Weinberg form factors. For the $F_1(q^2)$ and $G_1(q^2)$ form factors, we compare against QCD sum rule results from Ref.\cite{Zhang:2024ick}. While the couplings, i.e. the form factors at $q^2=0$, agree well, we observe tension in the $q^2$-dependence. 
In Fig.~\ref{fig:comp}, we compare commonly used coupling ratios with experimental data and phenomenological analyses. We find good agreement with the results of PDG and QCD sum rules, notably achieving a significantly higher precision. Our values are
{\thinmuskip=0mu
\medmuskip=1mu
\thickmuskip=2mu
\begin{equation}
    \frac{f_1^{\phantom{\text{S\!U\!(\!3\!)}}}}{f_1^\text{S\!U\!(\!3\!)}}{=}0.9674(47),\ \frac{g_1}{f_1}{=}0.6902(44),\ \frac{f_2}{f_1}{=}0.693(17),
\end{equation}}
where the error is the result of model averaging over different fit ranges to extract the matrix element and $f_1^{\text{S\!U\!(\!3\!)}} = \sqrt{3/2}$ is the $SU(3)$-flavor limit predicted by the Ademollo-Gatto theorem. 

\smallskip
\noindent
{\bf Determination of Decay Rates.}
We compute the decay rate of $\Lambda \to p\ell\bar{\nu}_\ell$ for both the electron ($\ell = e^-$) and muon ($\ell = \mu^-$) channels. Equation~\eqref{eq:decay_rate} is solved without fixing $|V_{us}|$, yielding results for $\Gamma/|V_{us}|^2$. As input, we use the leptonic masses $m_e = 0.511$\,MeV and $m_\mu = 0.10566$\,GeV, and the Fermi constant $G_F = 1.16638 \cdot (1 + 0.0105) \cdot 10^{-5}$\,GeV$^{-2}$\cite{ParticleDataGroup:2024cfk}, where the factor $(1 + 0.0105)$ accounts for radiative corrections~\cite{Garcia:1985xz}.
For the input on the baryon masses, we present two approaches:
(a) Using lattice-determined values, $m_N = 0.9471(28)$\,GeV and $m_\Lambda = 1.1263(23) $\,GeV, yields the results in Eqs.~(\ref{eq:res1},\ref{eq:res2}). Although the baryon mass uncertainties are at the permille level, the resulting mass splitting $m_\Lambda - m_N = 0.1792(24)$\,GeV carries a percent-level uncertainty. Since the decay rate scales with the fifth power of this splitting at leading order, the associated error is amplified by a factor of five, making it the dominant source of uncertainty.\\
(b) Using the experimental masses, $m^{\rm PRD}_p = 0.93827$\,GeV and $m^{\rm PRD}_\Lambda = 1.11568$\,GeV, eliminates this uncertainty, yielding the results in Eqs.(\ref{eq:res3}--\ref{eq:res5}). This choice illustrates the effect of using physical masses, however, on the current ensemble, we observe a percent-level discrepancy between lattice and experimental baryon masses—a known cutoff effect that vanishes in the continuum limit~\cite{Alexandrou:2023dlu}. This leads to a shift in the decay rate of approximately one standard deviation relative to approach (a).

We opt to quote as final results those obtained using the baryon masses determined from the lattice analysis (approach (a)), ensuring consistency with our simulation setup. We present results from approach (b) for comparison, to indicate the expected precision once the continuum-extrapolated form factors are combined with the physical masses.

\begin{figure}
\centering
{\includegraphics[width = \linewidth]{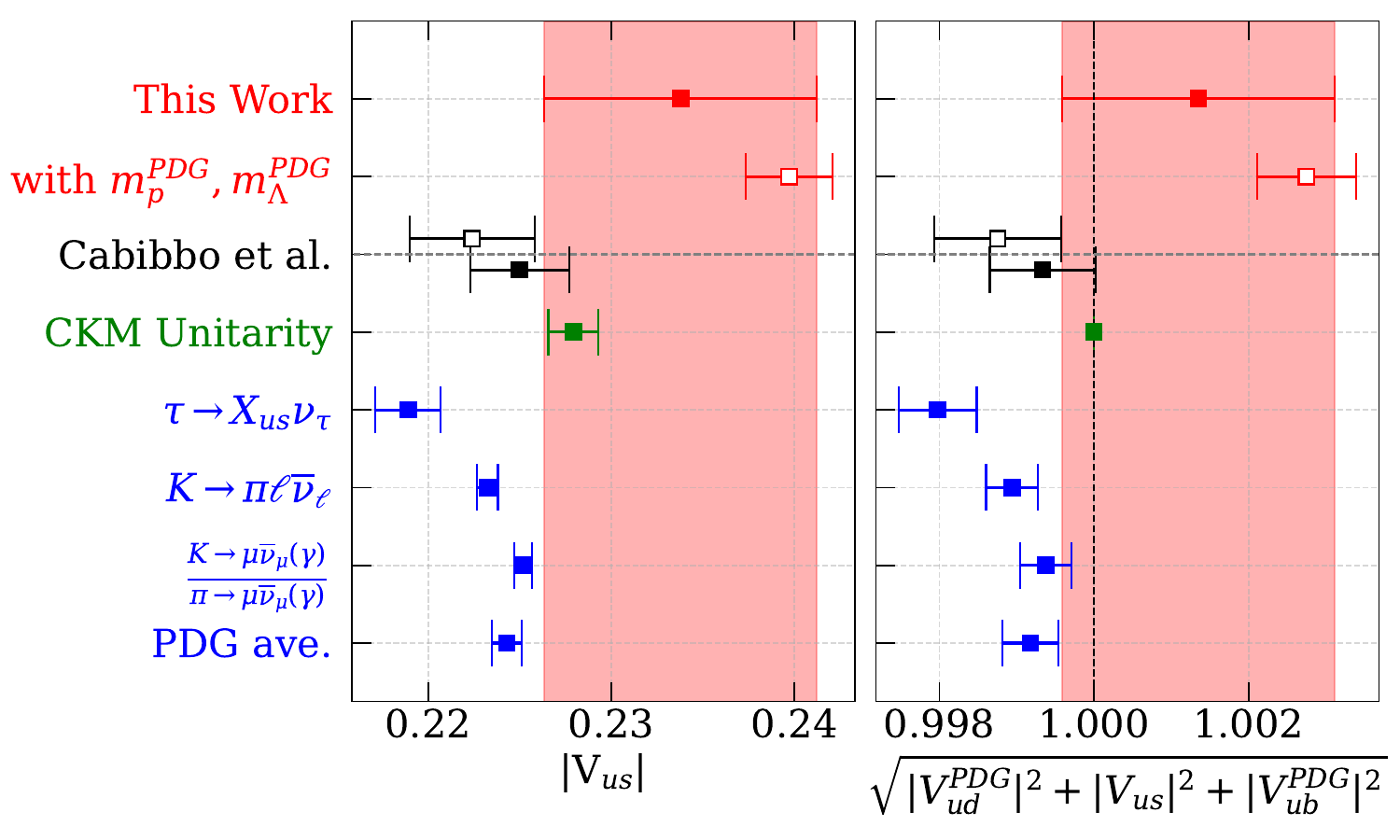}} 
\caption{Determination of $V_{us}$ (left) and the resulting CKM unitarity relation (right). Our results are shown in red: the top one uses lattice-determined nucleon and $\Lambda$ masses, approach (a), while the second one with open symbols uses PDG values, approach (b). The  black points correspond to the values by Cabbibo \textit{et al.}~\cite{Cabibbo:2003cu}, where the empty point is the value derived from the $\Lambda$ semileptonic decay, while the full point combines results from various hyperon decays. The green point is obtained from the unitarity relation. The remaining blue points show results on decay rates using lattice QCD combined with experimental data. Going from top to bottom, the  first blue square is from inclusive $\tau$ decays~\cite{ExtendedTwistedMass:2024myu}, the second from kaon semileptonic decays~\cite{Moulson:2017ive,Carrasco:2016kpy,Bazavov:2018kjg,FlavourLatticeAveragingGroupFLAG:2024oxs}, the third from the ratio of kaon to pion leptonic decays in the muonic channel~\cite{Dowdall:2013rya,Carrasco:2014poa,Bazavov:2017lyh,Miller:2020xhy,Alexandrou:2021bfr,Moulson:2017ive,FlavourLatticeAveragingGroupFLAG:2024oxs}, and the last is the PDG average of the latter two~\cite{ParticleDataGroup:2024cfk}. 
}
\label{fig:vues}
\end{figure}

\begin{figure}
\centering
{\includegraphics[width = 3.5in]{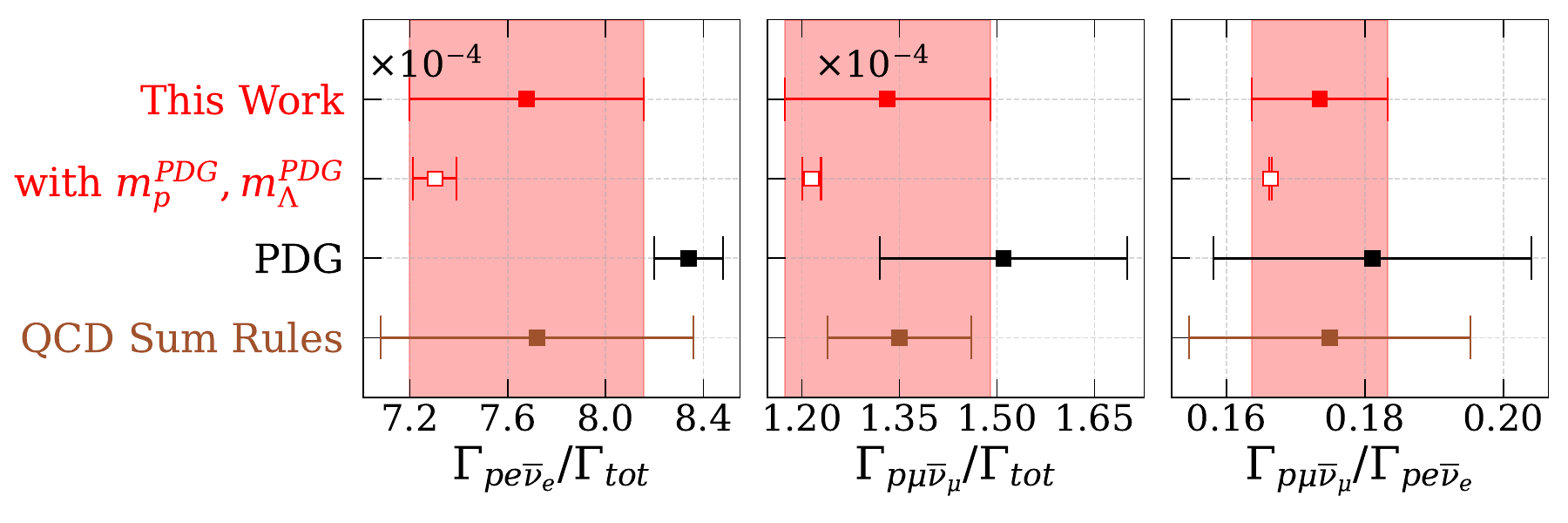}} 
\caption{Results on the decay rates in the electron and muon channels and their ratio. The notation of the red points is the same as in Fig.~\ref{fig:vues}. The PDG value is shown by the black points~\cite{ParticleDataGroup:2024cfk} and the QCD sum rules value by the brown points~\cite{Zhang:2024ick}. 
}
\label{fig:decrarte}
\end{figure}

\smallskip
\noindent
{\bf Final results.}
A robust, $|V_{us}|$-independent prediction from our analysis is the ratio of muon- to electron-channel decay rates,
\begin{align}
R^{\mu e} \equiv \frac{\Gamma_{p\mu\bar{\nu}_\mu}}{\Gamma_{pe\bar{\nu}_e}}  = 0.1735(98)\,.\label{eq:Rem}
\end{align}
To obtain further results, we consider two approaches:
(i) Using the experimental electron-channel decay rate, we extract $|V_{us}|$. Specifically, we take $\Gamma_{pe\bar{\nu}_e}^\text{PDG} = 8.34(14)\cdot 10^{-4}\times\Gamma_\text{tot}^\text{PDG}$ with the total decay rate $\Gamma_\text{tot}^\text{PDG} = 1/\tau_\Lambda^\text{PDG}$  and the $\Lambda$ lifetime $\tau_\Lambda^\text{PDG} = 2.617(10) \cdot 10^{-10}$\,s, yielding
\begin{align}
\big|V_{us}\big|\equiv \left(\frac{\Gamma_{ pe\bar{\nu}_e}}{\big|V_{us}\big|^2}\frac{1}{\Gamma_{ pe\bar{\nu}_e}^\text{PDG}}\right)^{-1/2} &= 0.2338(75)\,,\\
\sqrt{\big|V^\text{PDG}_{ud}\big|^2+\big|V_{us}\big|^2+\big|V^\text{PDG}_{ub}\big|^2} &=  1.0014(18)\,,
\end{align}
where the second relation tests first-row CKM unitarity, expected to equal unity in the Standard Model, using $|V_{ud}^{\rm PDG}| = 0.97367(32)$ and $|V_{ub}^{\rm PDG}| = 3.82(20)\cdot10^{-3}$~\cite{ParticleDataGroup:2024cfk}. 
 A comparison of these results with other determinations is shown in Fig.\ref{fig:vues}.
(ii) Alternatively, using as input the PDG average value $|V_{us}^{\rm PDG}| = 0.2243(8)$—extracted from semileptonic and leptonic Kaon decays— we obtain
\begin{align}
\frac{\Gamma_{ pe\bar{\nu}_e}}{\Gamma_\text{tot.}}&\equiv\frac{\Gamma_{ pe\bar{\nu}_e}}{\big|V_{us}\big|^2}\frac{\big|V_{us}^\text{PDG}\big|^2}{\Gamma_\text{tot.}^{\text{PDG}\phantom{^1}}}=7.68(48)\cdot 10^{-4},\label{eq:Vus}\\
\frac{\Gamma_{ p\mu\bar{\nu}_\mu}}{\Gamma_\text{tot.}}&\equiv\frac{\Gamma_{ p\mu\bar{\nu}_\mu}}{\big|V_{us}\big|^2}\frac{\big|V_{us}^\text{PDG}\big|^2}{\Gamma_\text{tot.}^{\text{PDG}\phantom{^1}}}=1.33(16)\cdot 10^{-4}\,.
\end{align}
A comparison of these results are shown in Fig.~\ref{fig:decrarte} including also the ratio of the two channels.

\section{Conclusions} \label{sec:conclusion}
We have presented the first lattice QCD calculation of the vector and axial-vector form factors governing the semileptonic decay $\Lambda \to p \ell \bar{\nu}_\ell$. These form factors are critical for probing the weak structure of hyperons and provide an independent avenue for determining the CKM matrix element $|V_{us}|$. Our calculation is performed on a single gauge-field ensemble with physical-mass up, down, strange, and charm quarks, eliminating the need for a chiral extrapolation. Discretization effects will be investigated in future studies but are expected to remain within the quoted uncertainties of our final results, based on corresponding results in the nucleon sector using the present ensemble and others at finer lattice spacings~\cite{Alexandrou:2025vto,Alexandrou:2023qbg}.

Our results for the form factor couplings are consistent with theoretical expectations, phenomenological models, and available experimental data. By integrating the differential decay rate over the full kinematic range, we obtain the partial decay rates for both electron and muon channels in units of $|V_{us}|^2$. This yields a precise lattice QCD prediction for their ratio, Eq.~\eqref{eq:Rem}, and a new determination of $|V_{us}|$, Eq.~\eqref{eq:Vus}, when the experimental decay rate of the electron channel is used as input.

The ratio computed from lattice QCD is significantly more precise than current experimental averages and is consistent with them, thereby ruling out the presence of scalar or tensor non-standard interactions at the current level of sensitivity~\cite{Chang:2014iba}. Our determination of $|V_{us}|$ is consistent with CKM unitarity, albeit with a larger uncertainty than other determinations that show a notable tension.
Compared to the phenomenological estimate from hyperon decays by Cabibbo \textit{et al.}~\cite{Cabibbo:2003cu}, we find that the approximations used in that analysis—such as substituting $f_1$ with $f_1^{\text{SU(3)}}$, neglecting second-class form factors, and ignoring the $q^2$ dependence—introduce systematic effects of a few percent. In particular, we find that neglecting the $q^2$-dependence leads to a 3.6(3)\% deviation in the electron-mode decay rate, while using the next-to-next-to-leading order in Eq.~\eqref{eq:cabibbo_2}, which incorporates $q^2$ dependence through the radii, results in a deviation of -1\%, as reported in Eq.~\eqref{eq:approx_diff}. This highlights the need for a non-perturbative approach, such as the present work, to achieve a sub-percent accuracy in the determination of $|V_{us}|$.

\section*{ACKNOWLEDGMENTS}
We thank Constantia Alexandrou for her insightful comments and mentorship as A.K.'s PhD advisor. We also thank the members of the ETM Collaboration for their constructive cooperation. S.B. acknowledge support from \texttt{POST-DOC/0524/0001} and \texttt{VISION ERC - PATH 2/0524/0001}, co-financed by the European Regional Development Fund and the Republic of Cyprus through the Research and Innovation Foundation within the framework of the Cohesion Policy Programme “THALIA 2021-2027”.. A.K. acknowledges financial support from ``The three-dimensional structure of the nucleon from lattice QCD'' \texttt{3D-N-LQCD} program, funded by the University of Cyprus.
Ensemble generation employed the open-source packages tmLQCD~\cite{Jansen:2009xp,Abdel-Rehim:2013wba,Deuzeman:2013xaa,Kostrzewa:2022hsv}, DD-$\alpha$AMG~\cite{Frommer:2013fsa,Alexandrou:2016izb,Bacchio:2017pcp,Alexandrou:2018wiv}, and QUDA~\cite{Clark:2009wm,Babich:2011np,Clark:2016rdz}. Analysis utilized the open-source software PLEGMA and QUDA. 
We gratefully acknowledge the Gauss Centre for Supercomputing e.V. (www.gauss-centre.eu) for computing time on JUWELS Booster at the Jülich Supercomputing Centre (JSC). We also acknowledge the Swiss National Supercomputing Centre (CSCS) and the EuroHPC Joint Undertaking for access to the Daint-Alps supercomputer. We are grateful to CINECA and the EuroHPC JU for access to the supercomputing facilities hosted at CINECA and Leonardo-Booster. 

\bibliography{main}{}

\appendix

\section{Appendix: END MATTER} \label{sec:end_matter}
We present additional details on the theoretical formalism and the extraction of final results. The generation and analysis of the lattice QCD data are discussed separately in the supplementary material.

\smallskip
\noindent
{\bf Helicity form factors.}
The helicity form factors are related to the Weinberg form factors via
\begin{align}
F_+(q^2) &= F_1(q^2) + \frac{q^2}{m_{\Lambda}(m_{\Lambda} + m_N)} F_2(q^2), \label{Eq:W3} \\
F_\perp(q^2) &= F_1(q^2) + \frac{m_{\Lambda} + m_N}{m_{\Lambda}} F_2(q^2), \label{Eq:W4} \\
F_0(q^2) &= F_1(q^2) + \frac{q^2}{m_{\Lambda}(m_{\Lambda} - m_N)} F_3(q^2), \label{Eq:W5}\\
G_+(q^2) &= G_1(q^2) - \frac{q^2}{m_{\Lambda}(m_{\Lambda} - m_N)} G_2(q^2), \label{Eq:W6} \\
G_\perp(q^2) &= G_1(q^2) - \frac{m_{\Lambda} - m_N}{m_{\Lambda}} G_2(q^2), \label{Eq:W7} \\
G_0(q^2) &= G_1(q^2) - \frac{q^2}{m_{\Lambda}(m_{\Lambda} + m_N)} G_3(q^2). \label{Eq:W8}
\end{align}
These expressions lead to the following decomposition of the matrix elements in Eq.~\eqref{eq:ME} for the vector $V_\mu = \bar{u}\gamma_\mu s$ and axial-vector $A_\mu = \bar{u}\gamma_\mu\gamma_5 s$ currents, respectively,
\label{appendix:Weinb}
\bgroup
\begin{widetext}
\thinmuskip=0mu
\medmuskip=1mu
\thickmuskip=2mu
\begin{align}
\label{eq:HFF}
M_\mu^\text{V} &= 
\bar{u}_p \Bigg[
    \frac{(m_{\Lambda} - m_p)q^\mu}{q^2}F_0(q^2)
    +  \frac{m_{\Lambda} + m_p}{s_+} \bigg(p^\mu_\Lambda + p^\mu_p - \frac{m_{\Lambda}^2 - m_p^2}{q^2} q^\mu \bigg)F_+(q^2) + \bigg( \gamma^\mu - \frac{2 m_Np^\mu_\Lambda}{s_+}  - \frac{2 m_{\Lambda}p^\mu_p}{s_+}  \bigg) F_\perp(q^2)
\Bigg] u_{\Lambda}\,,\\
M^\text{A}_\mu&= 
- \bar{u}_p \gamma_5 \Bigg[
     \frac{(m_{\Lambda} + m_N)q^\mu}{q^2} G_0(q^2)
    + \frac{m_{\Lambda} - m_N}{s_-} \bigg(p^\mu_\Lambda + p^\mu_p - \frac{m_{\Lambda}^2 - m_N^2}{q^2} q^\mu \bigg)G_+(q^2) + \bigg( \gamma^\mu + \frac{2 m_Np^\mu_\Lambda}{s_-}  - \frac{2 m_{\Lambda}p^\mu_p}{s_-}  \bigg) G_\perp(q^2)
\Bigg] u_{\Lambda}. \nonumber
\end{align}
\end{widetext}
\egroup
with $s_{\pm}=(m_\Lambda\pm m_N)^2-q^2$.
We also note that, at $q^2=0$, the helicity form factors satisfy 
\begin{equation}\label{eq:constraint1}
F_+(0){=}F_1(0){=}F_0(0)~~\text{and}~~G_+(0){=}G_1(0){=}G_0(0)\,,
\end{equation}
while, at the kinematic endpoint $q^2_{\max}=(m_\Lambda-m_N)^2$,
\begin{equation}\label{eq:constraint2}
 G_+(q^2_{\max}){=}G_1(q^2_{\max}){-}\frac{q^2_{\max}}{m_\Lambda}G_2(q^2_{\max}){=}G_\perp(q^2_{\max})   
\end{equation}
holds. These constraints are imposed in our parametrization of the form factors during the analysis.

\smallskip
\noindent
{\bf Parameterization of the $q^2$
 dependence.}\label{z-exp}
Since the relevant momentum transfer region for the $\Lambda \to p$ semileptonic decay lies entirely within the low-$q^2$ range, approximately within $[0,0.03]$\,GeV$^2$, a leading-order expansion in $q^2$ is sufficient to determine the decay rate with sub-percent precision. To this end, we define the axial-vector charges and radii as
\begin{equation}
    g_i = G_i(0)\quad\text{and}\quad
    \langle r^2_{G_i} \rangle = \frac{6}{g_i} \frac{\partial G_i(q^2)}{\partial q^2}\Bigg|_{q^2=0},
\end{equation}
and analogously for the vector form factors by substituting $G \leftrightarrow F$ and $g \leftrightarrow f$. These definitions yield the leading low-$q^2$ expansion
\begin{equation}
    G_i(q^2) = g_i \left(1+\frac{\langle r^2_{G_i} \rangle}{6} q^2\right) + \mathcal{O}(q^4)\,.\label{eq:lowq2}
\end{equation}
In contrast, the lattice data cover a significantly broader kinematic range, including negative $q^2$ and extending up to $q^2_{\max}$ for several form factors. To parametrize the full $q^2$ dependence in this regime, we adopt the model-independent $z$-expansion~\cite{Hill:2010yb}, as detailed in the supplementary material.

\smallskip
\noindent
{\bf Results on charges and radii.}
In Table~\ref{tab:chargeradi}, we provide our results for charges and radii in the Weinberg formalism, including correlation matrices between the parameters. For convenience, in the supplementary material, the same table can be found for the charges and radii of the helicity form factors.

\begin{table}[h]
    \centering
    \begin{tabular}{ccc}
        \hline
        \hline
        $f_1$ & $f_2$ & $f_3$ \\
        \hline
        1.1849(57) & 0.821(19) & 0.062(13)\\
        $\langle r^2_{F_1} \rangle m_\Lambda^2$ & $\langle r^2_{F_2} \rangle m_\Lambda^2$ & $\langle r^2_{F_3} \rangle m_\Lambda^2$ \\
        \hline
        9.50(28) & 17.0(23) & 24.7(68)\\
        \hline
        \hline
        $g_1$ & $g_2$ & $g_3$ \\
        \hline
        0.8178(55) & -0.082(19) & -5.14(11) \\
        $\langle r^2_{G_1} \rangle m_\Lambda^2$ & $\langle r^2_{G_2} \rangle m_\Lambda^2$ & $\langle r^2_{G_3} \rangle m_\Lambda^2$ \\
        \hline
        6.46(47) & 38(14) & 24.7(23)   \\        \hline
        \hline
    \end{tabular}

\begin{align*}
\begin{blockarray}{ccccccc}
~\text{corr.}~~ & f_1 & \langle r^2_{F_1} \rangle & f_2 & \langle r^2_{F_2} \rangle & f_3 & \langle r^2_{F_3} \rangle\\
\begin{block}{c(cccccc)}
f_1&1.00 & 0.11 & -0.03 & -0.12 & -0.50 & -0.33\\
\langle r^2_{F_1}\rangle&0.11 & 1.00 & -0.16 & -0.09 & -0.13 & -0.20 \\
f_2&-0.03 & -0.16 & 1.00 & 0.73 & 0.10 & 0.23 \\
\langle r^2_{F_2} \rangle&-0.12 & -0.09 & 0.73 & 1.00 & 0.20 & 0.36\\
f_3&-0.50 & -0.13 & 0.10 & 0.20 & 1.00 & 0.66\\
 \langle r^2_{F_3} \rangle&-0.33 & -0.20 & 0.23 & 0.36 & 0.66 & 1.00\\
\end{block}
\end{blockarray}\\
\begin{blockarray}{ccccccc}
~\text{corr.}~~ & g_1 & \langle r^2_{G_1} \rangle & g_2 & \langle r^2_{G_2} \rangle & g_3 & \langle r^2_{G_3} \rangle\\
\begin{block}{c(cccccc)}
g_1&1.00 & 0.26 & 0.50 & -0.05 & -0.19 & -0.02 \\
\langle r^2_{G_1} \rangle&0.26 & 1.00 & 0.47 & -0.44 & -0.01 & -0.30 \\
g_2&0.50 & 0.47 & 1.00 & -0.24 & -0.20 & -0.07  \\
\langle r^2_{G_2} \rangle&-0.05 & -0.44 & -0.24 & 1.00 & 0.20 & 0.80 \\
g_3&-0.19 & -0.01 & -0.20 & 0.20 & 1.00 & -0.15 \\
\langle r^2_{G_3} \rangle&-0.02 & -0.30 & -0.07 & 0.80 & -0.15 & 1.00 \\
\end{block}
\end{blockarray}
\end{align*}
    
    \caption{\label{tab:chargeradi}Charges and radii of the Weinberg form factors and the corresponding correlation matrix. The radii are provided in units of the experimental $m_\Lambda^2$ with $m_\Lambda = 1.11568$\,GeV.}
\end{table}

\begin{figure}
\centering
{\includegraphics[width = 3.5in]{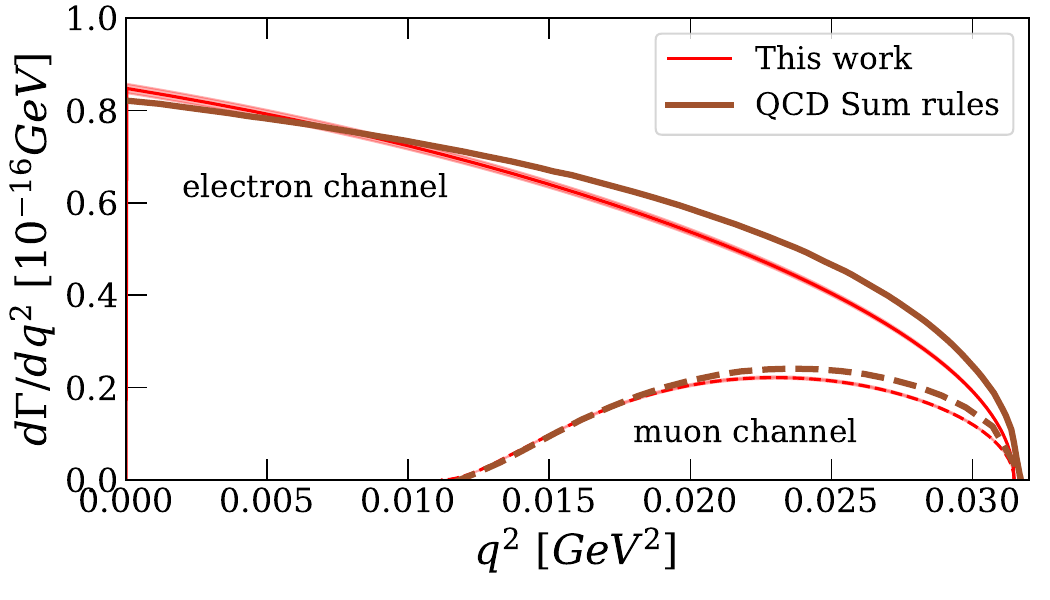}} 
\caption{Differential decay rates for the electron and muon channels. Brown lines show central values from the QCD sum rules study~\cite{Zhang:2024ick}, without error bands. 
In contrast, our results in red include uncertainty bands. 
The curves are computed using $|V_{us}^{\rm PDG}| = 0.2243(8)$ for visualization purposes.}
\label{fig:diff}
\end{figure}

\smallskip
\noindent
{\bf The differential decay rate.}
Using the helicity form factor parameterization of the hadronic matrix element yields the following expression for the differential decay rate~\cite{Meinel:2017ggx}
\begin{multline}
\frac{d\Gamma}{dq^2} =\frac{G_F^2 |V_{us}|^2 \sqrt{s_+ s_-}}{192 \pi^3 m_{\Lambda}^3}\left( 1 - \frac{m_\ell^2}{q^2} \right)^2 \times 
\\\Bigg\{\frac{3 m_\ell^2}{2q^2}\Big( s_- [(m_{\Lambda} + m_N) G_0]^2 + s_+ [(m_{\Lambda} - m_N) F_0]^2 \Big)+\\ \frac{m_\ell^2 + 2 q^2 }{2q^2}\Big( s_+ [(m_{\Lambda} - m_N) G_+]^2 + s_- [(m_{\Lambda} + m_N) F_+]^2 \Big)
\\+\left( m_\ell^2 + 2 q^2 \right)\Big( s_+ [G_\perp]^2 + s_- [F_\perp]^2 \Big)\Bigg\}\,
\label{eq:diff}
\end{multline}
which is integrated over the range $[m_\ell^2, q^2_{\max}]$ to obtain the semileptonic decay rate,
\begin{equation}
    \Gamma = \int_{m_\ell^2}^{q^2_{\max}} dq^2 \frac{d\Gamma}{dq^2}\,.\label{eq:decay}
\end{equation}
Figure~\ref{fig:diff} shows our determination of the differential decay rates for the electron and muon channels. For comparison, we include the QCD sum rules results from Ref.~\cite{Zhang:2024ick}. The deviation observed is likely driven by the discrepancy in the $q^2$-dependence, already apparent in Fig.~\ref{fig:WFF}. 

When integrating Eq.~\ref{eq:diff}, all input parameters are precisely known except for the CKM matrix element $|V_{us}|$, which exhibits a few-percent-level tension among different determinations. As a result, we can provide a robust estimate only for the ratio $\Gamma/|V_{us}|^2$, computed for both the electron and muon channels, obtaining
\begin{align}\label{eq:res1}
    \frac{\Gamma_{pe\bar{\nu}_e}}{|V_{us}|^2} &= 5.83(37)\cdot10^{7} \,{\rm s}^{-1}\,,\\\label{eq:res2}
    \frac{\Gamma_{p\mu\bar{\nu}_\mu}}{|V_{us}|^2} &=1.01(12)\cdot10^{7} \,{\rm s}^{-1}\,.
\end{align}
Using the experimental masses instead, we obtain
\begin{align}\label{eq:res3}
    \frac{\Gamma_{pe\bar{\nu}_e}}{|V_{us}|^2} &= 5.546(51)\cdot10^{7}\,{\rm s}^{-1}\,,\\\label{eq:res4}
    \frac{\Gamma_{p\mu\bar{\nu}_\mu}}{|V_{us}|^2} &=0.9228(82)\cdot10^{7}\,{\rm s}^{-1}\,,\\\label{eq:res5}
        \frac{\Gamma_{p\mu\bar{\nu}_\mu}}{\Gamma_{pe\bar{\nu}_e}} &=0.16638(20).
\end{align}
Notably, in the latter result, the uncertainties are significantly reduced due to strong correlations between the two decay rate determinations, leading to a substantial cancellation of statistical errors.

\smallskip
\noindent
{\bf Perturbative approximation of the decay rate.}
Given the narrow kinematic range, the decay rate can be computed as a perturbative expansion in the small dimensionless parameter $\delta=q_{\max}/m_\Lambda\approx 0.16$ and using the low-$q^2$ approximation of the form factors in Eq.~\eqref{eq:lowq2}.  Restricting to the electron channel, $\Lambda \to pe\bar{\nu}_e$, further simplifies the calculation by allowing the lepton mass to be neglected. Under these approximations, the next-to-leading order (NLO) expression for the decay rate is~\cite{Cabibbo:2003cu}
\begin{multline}\label{eq:cabibbo}
    \Gamma_{\rm NLO} = \frac{G_F^2 |V_{us}|^2  m_\Lambda^5\delta^5}{60 \pi^3}\left(1-\frac{3}{2}\delta\right)\bigg(f_1^2+3g_1^2\bigg)\,.
\end{multline}
At next-to-next-to-leading order (NNLO), we find
\begin{align}\label{eq:cabibbo_2}
&\!\!\!\Gamma_{\rm NNLO}=\Gamma_{\rm NLO}+\frac{G_F^2 |V_{us}|^2 m_\Lambda^5 \delta ^7 }{70 \pi ^3} \Bigg\{ f_1f_2 +\frac{2}{3} f_2^2-\frac{14}{3}\frac{g_1g_2}{\delta}\nonumber\\ &\quad+  f_1^2  \bigg(1+\frac{1}{9} \langle r^2_{F_1} \rangle m_\Lambda^2\bigg)+ g_1^2  \bigg(2+\frac{5}{9} \langle r^2_{G_1} \rangle m_\Lambda^2\bigg) \Bigg\}\,
\end{align}
where the $g_1 g_2/\delta$ term contributes at this order since $g_2$ is a second-class form factor with $g_2 \propto \delta$. Compared to the NNLO corrections given in Eq.~(40) of Ref.~\cite{Cabibbo:2003cu}, our result reinstates essential contributions proportional to the radii, arising from the $q^2$-dependence of the $F_1(q^2)$ and $G_1(q^2)$ form factors, which were previously omitted. When comparing to the full non-perturbative result, we find
\begin{equation}\label{eq:approx_diff}
\Delta_\text{NLO} = 5.5(2)\% \quad\text{and}\quad \Delta_\text{NNLO} = -1.14(8)\%,
\end{equation}
having defined $\Gamma = \Gamma_\text{(N)NLO}(1 + \Delta_\text{(N)NLO})$.

\clearpage

\section{SUPPLEMENTARY MATERIAL} \label{appendix}

The gauge ensemble analyzed is generated using the twisted-mass fermion discretization scheme, which provides automatic $\mathcal{O}(a)$-improvement~\cite{Frezzotti:2000nk,Frezzotti:2003ni}. A clover term is included in the action~\cite{Sheikholeslami:1985ij} to suppress the isospin-breaking effects inherent to this fermion discretization. The isosymmetric pion mass is tuned to $ m_\pi = 135 $ MeV~\cite{Alexandrou:2018egz,Finkenrath:2022eon} by adjusting the bare light quark mass parameter $ \mu_l $. The strange and charm quark mass parameters, $ \mu_s $ and $ \mu_c $, are determined using the kaon mass and a well-defined ratio of the D-meson mass to its decay constant, as well as a phenomenologically motivated ratio between the strange and charm quark masses, following the approach of Ref.~\cite{Alexandrou:2018egz,Finkenrath:2022eon}. The parameters of the ensemble used in this analysis are summarized in Table~\ref{tbl:Ensembles}. 
The determination of the lattice spacing used in this work is the one extracted in Ref.~\cite{ExtendedTwistedMass:2022jpw}. We note that,  although lattice spacings obtained using  different quantities may exhibit slight differences due to cutoff effects, the results in the continuum limit remain fully consistent when parameters are determined in a self-consistent manner~\cite{ExtendedTwistedMass:2021gbo}. 
\begin{table}[h!]
	\centering
	{\small
		\renewcommand{\arraystretch}{1.2}
		\renewcommand{\tabcolsep}{1.5pt}
    \begin{tabular}{c|c|c|c|c|c}
    \hline \hline
        Ensemble     & Abrv. & $V/a^4$            & $\beta$ & $a$~[fm]    & $m_\pi$~[MeV] \\ \hline 
        cB211.072.64 & B64   & $64^3 \times 128$     & 1.778   & 0.07957(13) & 140.2(2)      \\ 
         \hline \hline
        \end{tabular}}
    \caption{Parameters of the $N_f=2+1+1 $ ensemble analyzed in this work. In the first column, we give the name of
             the ensemble, in the second the abbreviated name, in the third the lattice volume, in the fourth $\beta = 6/g^2$ with $g$ the bare coupling constant, in the fifth
             the lattice spacing and in the last the pion mass. The lattice spacing and the pion mass is determined in Ref.~\cite{ExtendedTwistedMass:2022jpw}.}
    \label{tbl:Ensembles}
\end{table}

\subsection{Generation of correlation functions}

The necessary correlation functions for extracting matrix elements of the $\Lambda \to N$ transition are produced at the simulated light quark mass $a\mu_{ud, \text{sea}} = 0.00072$ and a strange quark mass $a\mu_{s,\text{isoQCD}} = 0.018267(53)$, fine-tuned~\cite{ExtendedTwistedMass:2024nyi} to match our target definition of isoQCD as defined by the Edinburgh/FLAG consensus~\cite{FlavourLatticeAveragingGroupFLAG:2024oxs}. Corrections due to the small mistuning of the light quark mass $a\mu_{ud,\text{isoQCD}} = 0.0006669(28)$ are expected to lie below the percent level, as estimated in related studies for other quantities~\cite{ExtendedTwistedMass:2022jpw,ExtendedTwistedMass:2024nyi,Alexandrou:2023dlu},  and therefore below the targeted precision.

In addition, matrix elements computed in lattice QCD require renormalization to relate them to physical observables. In this work, we employ non-perturbative renormalization techniques based on Ward identities and the universality of renormalized hadronic matrix elements, commonly referred to as hadronic renormalization methods. These methods do not require gauge fixing, unlike the RI$'$ scheme, and provide higher precision. For the axial isovector current renormalization factor $ Z_A $, we find 
\begin{equation}
    Z_A = 0.74294(24) \quad \text{for cB211.072.64~\cite{ExtendedTwistedMass:2022jpw}\,,}
\end{equation}
which renormalizes  multiplicatively  matrix elements produced with the local axial-vector current operator $A_\mu(x_{\rm ins})=\bar{u}(x)\gamma_\mu\gamma_5 s(x)$.
For the vector current, no renormalization is needed since we use the conserved current operator,
\begin{multline}
V_\mu^{\rm cons.}(x) = \bar{u}(x)\Big[(1 + \gamma_\mu)(\cev{T}_{+\mu} + \vec{T}_{-\mu})+\\ - (1 - \gamma_\mu)(\cev{T}_{-\mu} + \vec{T}_{+\mu})\Big]s(x),
\end{multline}
where $\vec{T}_\mu$ and $\cev{T}_\mu$ are the forward and backward gauge-covariant transport operators that act on the spinor fields as $\vec{T}_\mu \psi(x) = U_\mu(x)\psi(x + \hat{\mu})$ and 
 $\bar{\psi}(x) \cev{T}_\mu = \bar{\psi}(x + \hat{\mu}) U^\dagger_\mu(x).
$

\subsection{Smearing and statistics}

Correlation functions suffer from an exponentially degrading signal-to-noise ratio (SNR) at large Euclidean time separations. To mitigate this, we compute correlation functions using smeared point sources at multiple source positions per configuration, shift all sources to a common reference point ($x_0 = 0$), and average over them. Smearing enhances the overlap of interpolating fields with the ground state. We employ Gaussian smearing~\cite{Gusken:1989ad,Alexandrou:1992ti} in the spatial directions, modifying the quark fields as
\begin{equation}
q_G(\vec{x}, t) = \left[ 1 + \alpha_G \sum_{j=1}^3 \left(\vec{T}_{+j} + \vec{T}_{-j} \right) \right]^{N_G} q(\vec{x}, t),
\end{equation}
where $\alpha_G$ is the smearing parameter, $N_G$ the number of iterations, and $j=1,2,3$ denotes spatial directions. Gauge links in the smearing operator are APE-smeared~\cite{APE:1987ehd}. The number of Gaussian smearing steps differs between strange and light quarks, as detailed in Table~\ref{tab:stat}.

In Table~\ref{tab:stat}, we also give the number of source positions used for each source-sink separation in the three-point function as well as  for the nucleon and $\Lambda$ two-point functions. For three-point functions, statistics are typically increased exponentially with the source-sink separation to keep statistical errors approximately constant at least up to $t_{\rm s} = 1.12$~fm. Beyond this, due to computational cost, statistics are held fixed, resulting in exponentially growing errors. These longer separations contribute less to the final statistical precision but remain valuable for controlling excited-state contamination. The increased number of source positions for the nucleon two-point functions is inherited from other nucleon structure studies using the same ensemble.

\begin{table}[h!]
\centering
\renewcommand{\arraystretch}{1}
\setlength{\tabcolsep}{3pt}

    \begin{tabular}{|l|c|c|c|c|}
    \hline
\multicolumn{5}{|c|}{\textbf{Smearing parameters}} \\
        \hline
        Flavor& $a_G$ & $n_G$ & $a_\text{APE}$ & $n_\text{APE}$\\
        \hline
        Light & 1.0 & 95 & 0.5& 50\\
        Strange & 1.0 & 40 & 0.5& 50\\
        \hline
    \end{tabular}~~
\begin{tabular}{|c|c|c|}
\hline
\multicolumn{3}{|c|}{\textbf{cB211.072.64}} \\
\multicolumn{3}{|c|}{524 configurations} \\
\hline
$t_{\rm s}/a$ & $t_{\rm s}$ [fm] & $n_{src}$ \\
\hline
8 & 0.64 & 1 \\
10 & 0.80 & 3 \\
12 & 0.95 & 9 \\
14 & 1.12 & 27 \\
16 & 1.28 & 27 \\
18 & 1.44 & 27 \\
\hline
\multicolumn{2}{|l|}{Nucleon 2pt} & 349 \\
\multicolumn{2}{|l|}{$\Lambda$-hyperon 2pt} & 94 \\
\hline
\end{tabular}

\caption{Left: Smearing parameters for the light and strange quark interpolating fields. Right: Statistics for the three-point functions at various source-sink separations, along with the total statistics used for the nucleon and $\Lambda$ two-point functions in the analysis.}
\label{tab:stat}
\end{table}

\subsection{Ground state extraction}

\begin{figure}
\centering
\includegraphics[width = \linewidth]{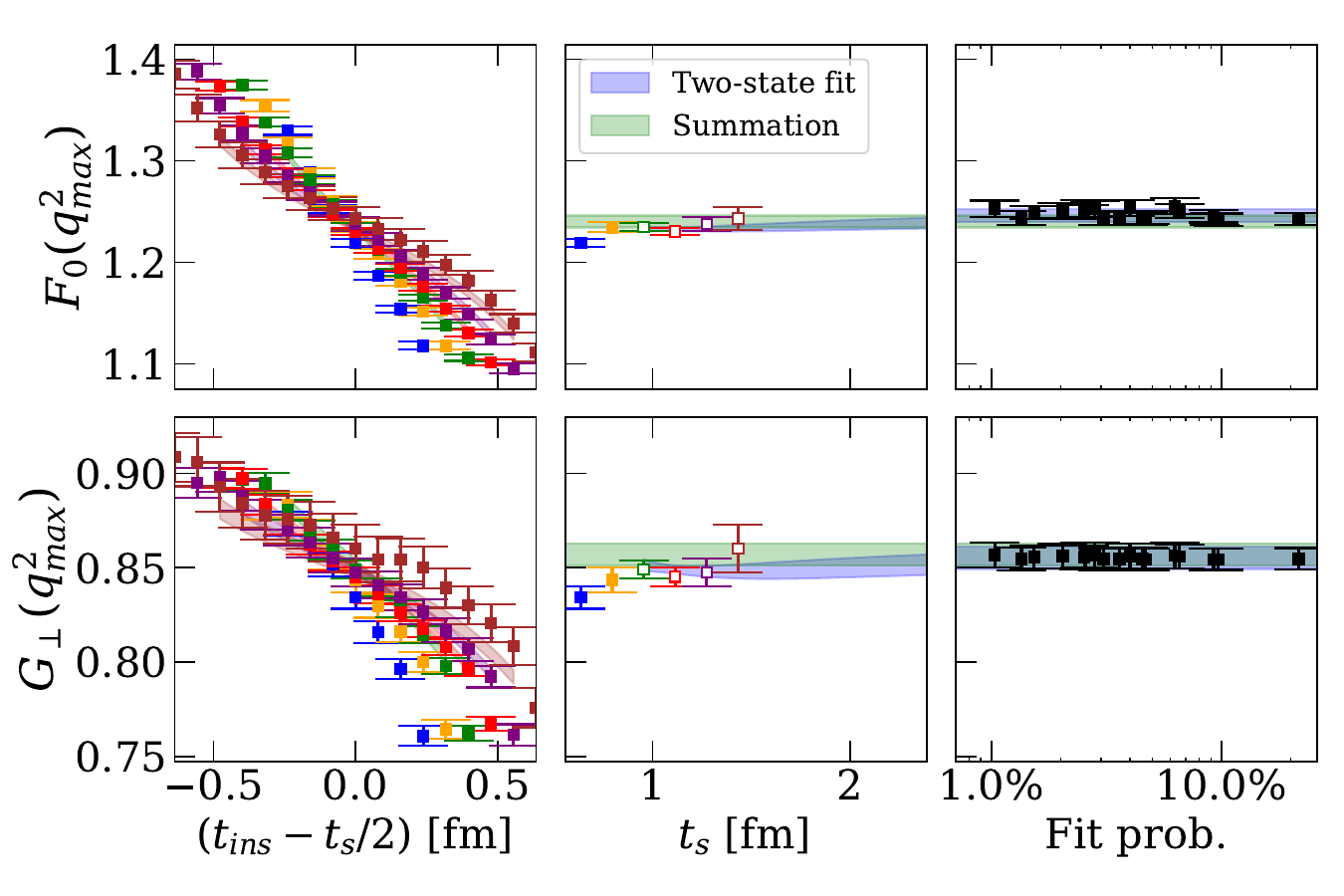}
\caption{Results on the ratios of three- to two-point correlation functions, defined in Eq.~\eqref{eq:ratio}, are shown for the kinematic point $\vec{p}_N = \vec{p}_\Lambda = \vec{0}$, where they asymptotically yield the form factors $F_0(q^2)$ and $G_\perp(q^2)$ at $q^2_{\rm max}$ after proper normalization and renormalization. The left panels display the $t_{\rm ins}$ dependence for the various source-sink separations $t_{\rm s}$ and the middle panels show the $t_{\rm s}$ dependence of the midpoint $t_{\rm ins} = t_{\rm s}/2$, with matched colors across all panels to indicate corresponding $t_{\rm s}$ values. The right panels present the extracted asymptotic values from the two-state analysis, plotted against fit probabilities greater than 1\%. The green bands indicate the model-averaged results from the summation method, while the blue bands represent the model-averaged  results of the two-sate fits. For clarity, all form factors are rescaled by $e^{-(m_\Lambda - m_N)(t_{\rm ins} - t_{\rm s}/2)}$ to account for the residual $t_{\rm ins}$ dependence of the ground-state.}
\label{fig:fitFFs}
\end{figure}

Controlling excited-state contamination is critical for reliably extracting ground-state matrix elements. In this work, we follow state-of-the-art procedures developed in the nucleon sector~\cite{Alexandrou:2024ozj,Alexandrou:2023qbg,Alexandrou:2025vto}. Such analyses are particularly important for nucleon matrix elements of the axial current, where contamination is enhanced by large overlaps with $\pi N$ states~\cite{Bar:2018xyi,Jang:2019vkm}. In the matrix elements studied here, the analogous enhancement comes from $KN$ states, resulting in an energy gap between $\Lambda$ and $KN$ roughly twice as large as that between $N$ and $\pi N$, and therefore much more exponentially suppressed. This is confirmed by the convergence of the extracted matrix elements shown in Fig.~\ref{fig:fitFFs}. We further mitigate even this small contamination by performing a thorough study of excited-state effects, as outlined below.

\begin{figure}
\centering
{\includegraphics[width = \linewidth]{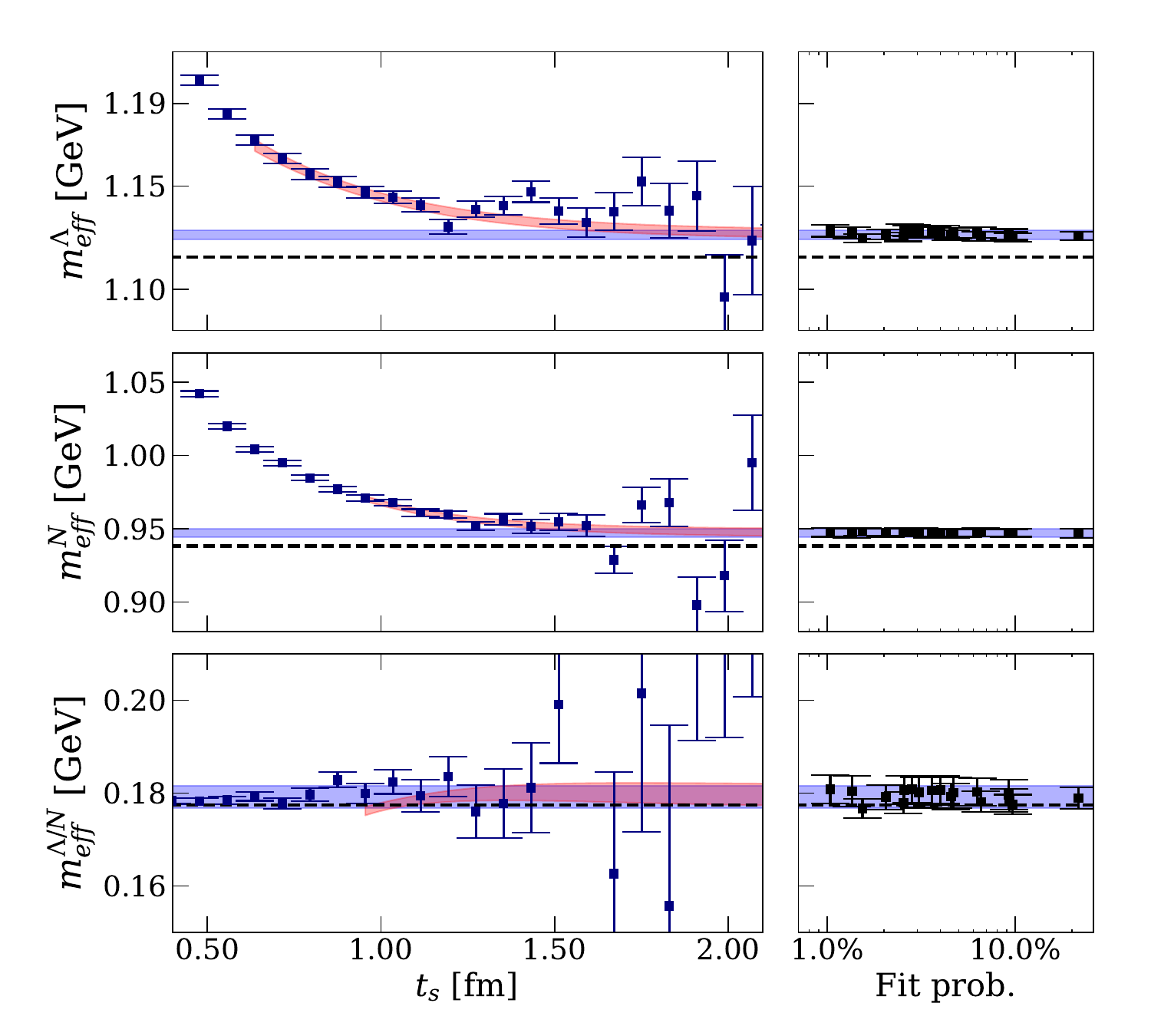}}

\caption{Results for the effective masses for the $\Lambda$ hyperon, nucleon, and their mass splitting as functions of the source-sink separation $t_{\rm s}$ are shown in the left panels, while the right panels display results from two-state fits with model probabilities exceeding 1\%. Horizontal dashed lines indicate the experimental values, and the blue bands represent model-averaged results from all accepted two-state fits, yielding $m_\Lambda = 1.1263(23)$\,GeV, $m_N = 0.9471(28)$\,GeV, and $m_\Lambda - m_N = 0.1792(24)$\,GeV. Red bands denote the most probable fits with a reduced $\chi^2$ of $\chi^2/N_{\rm dof} = 1.51$. The ground-state masses are determined in a simultaneous fit of the three effective masses and the three-point function ratios at $\vec{p}_N = \vec{p}_\Lambda = \vec{0}$ shown in Fig.~\ref{fig:fitFFs}.}
\label{fig:Mass}
\end{figure}

We expand the correlation functions in terms of its eigenspectrum and limit  contributions up to the  first excited state. The two-point function, Eq.~\eqref{eq:2pt}, is then given as
\begin{equation}
C_{B}(\vec{p},t_{\rm s}) = e^{-t_{\rm s}E_B^{0|\vec{p}|}}(c_B^{0|\vec{p}|}+c_B^{1|\vec{p}|}e^{-t_{\rm s}\Delta E_B^{1|\vec{p}|}}),
\label{two-state}
\end{equation}
where $E_B^{0|\vec{p}|}$ is the ground-state energy, and $\Delta E_B^{1|\vec{p}|} = E_B^{1|\vec{p}|} - E_B^{0|\vec{p}|}$ is the energy gap to the first excited state. The amplitudes
\begin{equation}
c_B^{i|\vec{p}|} = \Tr\Big[P_0\langle \chi_B(\vec{p})|B_i(\vec{p})\rangle\langle B_i(\vec{p})|\chi_B(\vec{p})\rangle\,\Big]
\end{equation}
---with the trace taken over the implicit Dirac indices of the interpolating fields---are proportional to the squared overlap of the interpolating field with the $i$-th energy eigenstate. Due to lattice rotational symmetry, these are independent of the direction of $\vec{p}$.
The corresponding expansion of the three-point function, Eq.~\eqref{eq:3pt}, yields
{\thinmuskip=0mu
\medmuskip=1mu
\thickmuskip=2mu
\begin{align}
&C^A_{\mu\nu}(\vec{p}_N,\vec{p}_\Lambda;t_{\rm s},t_{\rm ins})=
e^{-(t_{\rm s}-t_{\rm ins})E_\Lambda^{0|\vec{p}_\Lambda|}-t_{\rm ins}E_N^{0|\vec{p}_N|}}\Big(A^{00}_{\mu\nu}(\vec{p}_N,\vec{p}_\Lambda)\nonumber\\
&~~+A^{01}_{\mu\nu}(\vec{p}_N,\vec{p}_\Lambda)e^{-t_{\rm ins}\Delta E_N^{1|\vec{p}_N|}}+
A^{10}_{\mu\nu}(\vec{p}_N,\vec{p}_\Lambda)e^{-(t_{\rm s}-t_{\rm ins})\Delta E_\Lambda^{1|\vec{p}_\Lambda|}}\nonumber\\
&~~+A^{11}_{\mu\nu}(\vec{p}_N,\vec{p}_\Lambda)e^{-(t_{\rm s}-t_{\rm ins})\Delta E_\Lambda^{1|\vec{p}_\Lambda|}-t_{\rm ins}\Delta E_N^{1|\vec{p}_N|}}\Big)\,,
\end{align}
}
where each amplitude $A^{ij}_{\mu\nu}$ contains the product of interpolator overlaps and matrix elements, given by
\begin{multline}
A^{ij}_{\mu\nu}(\vec{p}_N,\vec{p}_\Lambda)=\Tr\Big[P_\nu\langle \chi_\Lambda(\vec{p}_\Lambda)|\Lambda_i(\vec{p}_\Lambda)\rangle\times\\\langle\Lambda_i(\vec{p}_\Lambda)| \bar{u}\varGamma^\text{A}_\mu s|N_j(\vec{p}_N)\rangle\langle N_j(\vec{p}_N)|\chi_N(\vec{p}_N)\rangle \Big].
\end{multline}
A similar expansion applies to the vector-current three-point functions, with $A \leftrightarrow V$.
The desired ground-state matrix elements, defined in Eq.~\eqref{eq:LME}, are obtained from
\begin{equation}
    \Pi^\text{A}_{\mu\nu}(\vec{p}_N,\vec{p}_\Lambda) = \frac{A^{00}_{\mu\nu}(\vec{p}_N,\vec{p}_\Lambda)}{\sqrt{c^0_\Lambda(\vec{p}_\Lambda)c^0_N(\vec{p}_N)}}\,,
\end{equation}
where the overlap factors cancel, isolating the desired matrix element.

\begin{figure}
\centering
{\includegraphics[width = \linewidth]{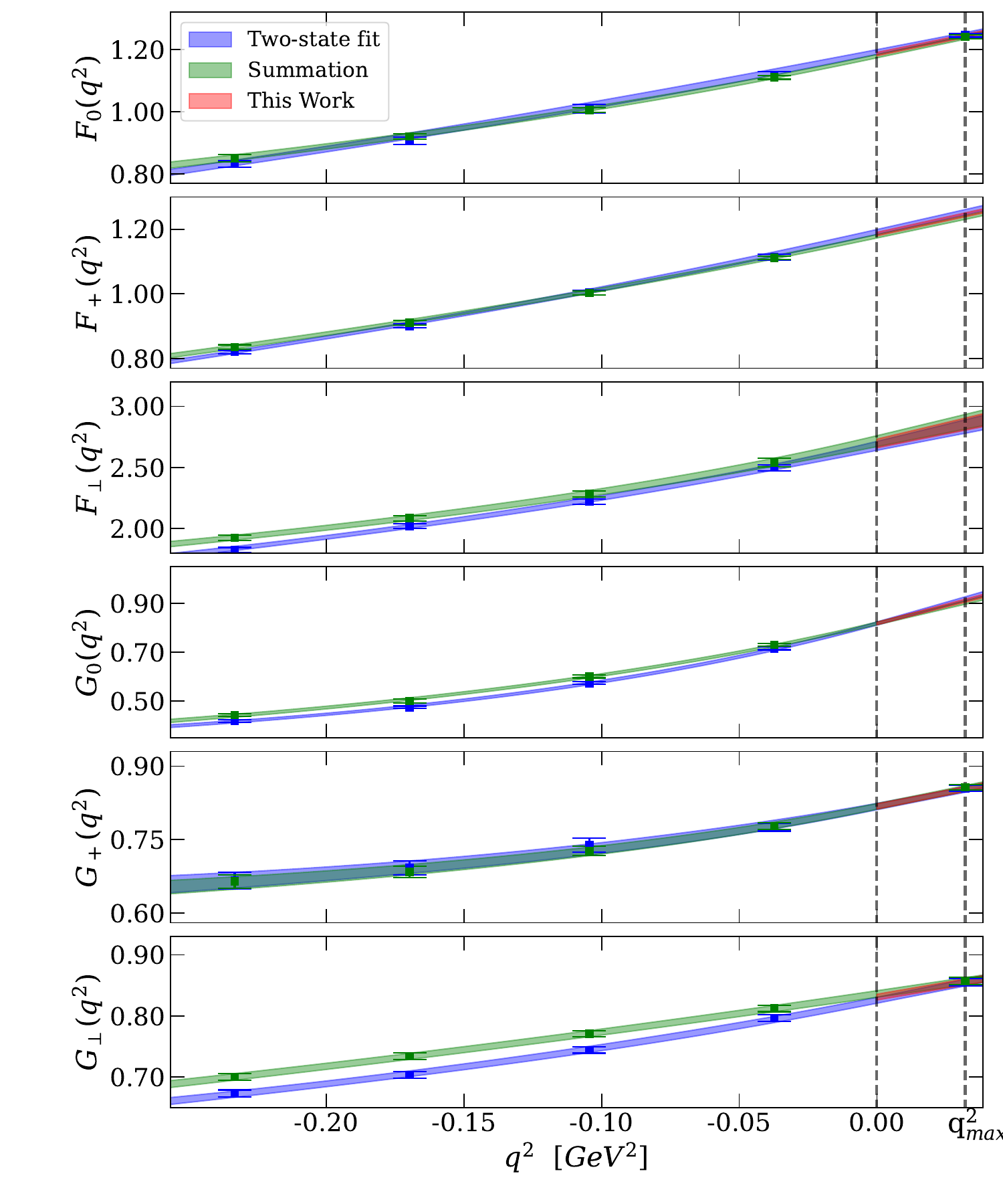}}
\caption{Momentum-transfer dependence of the helicity form factors. Blue points show results from the two-state fit, and green points from the summation method. The corresponding $z$-expansion fits are shown as blue and green bands, respectively. The red band represents the final combined result. Vertical dashed lines indicate the kinematic region relevant for the semileptonic decay, $q^2 \in [0, q^2_{\max}]$.}
\label{fig:HFF}
\end{figure}

\subsection{Fitting strategy}

Based on the above two-state expansion, to accurately extract ground-state matrix elements and access contributions from excited-states, we employ two complementary methods: the summation and the two-state fit methods. 

In the \textit{summation method}, we construct a modified ratio with respect to Eq.~\eqref{eq:ratio} designed to fully cancel the ground-state time dependence, including its dependence on the operator insertion time $t_{\rm ins}$~\cite{Alexandrou:2013joa,Alexandrou:2011db,Alexandrou:2006ru}. Summing over $t_{\rm ins}$ in a range $[n,t_{\rm s}-n]$ yields
\begin{multline}
\label{eq:summeth}
S^{\rm A}_{\mu\nu}
(\vec{p}_N,\vec{p}_\Lambda;t_{\rm s})\equiv \sum_{t_{\rm ins}=n}^{t_{\rm s}-n}R^{\rm A}_{\mu\nu}
(\vec{p}_N,\vec{p}_\Lambda;t_{\rm s},t_{\rm s})\times\\ \frac{\sqrt{C_\Lambda(\vec{p}_\Lambda;t_{\rm ins})
 C_N(\vec{p}_N;t_{\rm s}-t_{\rm ins})}}{\sqrt{C_N(\vec{p}_N;t_{\rm ins})C_\Lambda(\vec{p}_\Lambda,t_{\rm s}-t_{\rm ins})
}}
\end{multline}
which leads to a geometric series over excited-state contributions~\cite{Maiani:1987by,Capitani:2012gj}. The resulting summed ratio behaves as
\begin{equation}
S^{\rm A}_{\mu\nu}
(\vec{p}_N,\vec{p}_\Lambda;t_{\rm s})= c + \Pi_{\mu\nu}^{\rm A}(\vec{p}_N,\vec{p}_\Lambda)\, t_{s} + \dots    
\end{equation}
where the ground-state matrix element appears as the slope of a linear fit in $t_{\rm s}$. The omitted terms given by the eclipse are dominated by excited-state contamination and decay exponentially with $t_{\rm s}$. Compared to the standard ratio method, this approach achieves a parametrically faster suppression of excited-state effects, making it particularly effective if only one-state is kept. Another advantage of the summation method is its reduced fitting complexity, involving only two hyperparameters: the summation range, controlled by $n$, and the minimum source-sink separation, $t_{\rm s,min}$, from which the linear fit is performed. To address potential systematic uncertainties arising from different fit configurations, we use a model-averaging approach, also known as the Akaike Information Criterion (AIC)~\cite{Jay:2020jkz,Neil:2022joj} to remove the dependence on these hyperparameters.  Briefly, for each fit $ i $, we assign a weight $ w_i $, defined as
\begin{equation}\label{eq:weight}
\log(w_i) = -\frac{\chi_i^2}{2} + N_{\text{dof}, i},
\end{equation}
where $ N_{\text{dof}, i} = N_{\text{data}, i} - N_{\text{params}, i} $ is the number of degrees of freedom, calculated as the difference between the number of data points $ N_{\text{data}, i} $ and the number of fitting parameters $ N_{\text{params}, i} $ and $ \chi^2_i $ is the chi-squared value of the correlated fit.

In the \textit{two-state fit},  we extract the energies and amplitudes from the two-state expansion of the correlation functions. Rather than fitting the correlators themselves—which would involve a larger set of correlated parameters—we fit appropriately constructed ratios. This approach reduces statistical correlations between data points and  the number of fit parameters eliminating the ground-state overlaps.
For the two-point functions, we use the effective-energy ratio
\begin{equation}
    \label{eq:fit1}
E_B^{\rm eff}(t_{\rm s},\vec{p}) = \log\frac{C_B(t_{\rm s},\vec{p})}{C_B(t_{\rm s}+1,\vec{p})}= E^N_0(\vec{p})+\dots,
\end{equation}
which asymptotically approaches the ground-state energy. We perform two-state fits to the $t_{\rm s}$ behavior using the expansion given in Eq.~(\ref{two-state}).
For the three-point functions, we use the ratio in Eq.\eqref{eq:ratio}, which isolates the matrix elements in the large-time limit as described in Eq.\eqref{eq:LME}.
To further improve the determination of the $\Lambda$–$N$ mass splitting, we construct a ratio of two-point functions at zero momentum using shared statistics
\begin{equation}
\label{eq:fit2m}
m^{\Lambda/N}_{\rm eff}(t_{\rm s}) = \log\frac{C^{\Lambda}(t_{\rm s})C^{N}(t_{\rm s}+1)}{C^{N}(t_{\rm s})C^{\Lambda}(t_{\rm s}+1)}=m_\Lambda-m_N+\dots,
\end{equation}
which directly isolates the mass difference in the asymptotic limit. Since the energies are common between two- and three-point functions, we perform a combined fit of these ratios at each unit of momentum transfer. In these fits, we employ five hyperparameters: the initial fit time for the $\Lambda$ effective energy, $t_{{\rm s,min,\Lambda}}$; for the nucleon effective energy, $t_{{\rm s,min,N}}$; for the three-point functions with respect to the source-sink separation, $t_{{\rm s,min,3pt}}$; the minimum insertion time on the source side, $t_{{\rm ins,min,3pt}}$; and the maximum insertion time on the sink side, $t_{{\rm ins,max,3pt}}$. These hyper-parameters, are varied independently within reasonable ranges to optimize the region in which the two-state dominate in the correlation functions. As in the case of the summation method, the dependence on the hyper-parameters is removed using the model average procedure with weights defined in Eq.~\eqref{eq:weight}.

The results of this procedure are shown in Fig.~\ref{fig:Mass} for the effective masses and in Fig.~\ref{fig:fitFFs} for the form factors evaluated at $q^2_{\max}$, corresponding to the kinematic point $\vec{p}_N = \vec{p}_\Lambda = \vec{0}$. The same analysis is repeated for all accessible momentum units of $\vec{p}_N$, while keeping $\vec{p}_\Lambda = \vec{0}$ fixed for all three-point functions, as dictated by our fixed-sink setup.

\subsection{Decomposition of matrix elements}

The final step of the analysis involves extracting the form factors from the ground-state matrix elements. For our setup with $\vec{p}_\Lambda = \vec{0}$, the relevant non-zero components in the helicity formalism used to extract form factors at the respective $q^2$ are given by

\begin{widetext}
{\thinmuskip=0mu
\medmuskip=1mu
\thickmuskip=2mu
\begin{align}
\Pi^V_{ij}(\vec{p}_N)=\,&\frac{\epsilon_{ijk}p_N^k {\cal K}}{2m_N}\,F_\perp(q^2)\,,\\
\Pi^V_{0 0}(\vec{p}_N)=\,&
\frac{(E_N+m_N){\cal K}}{2m_N} \bigg\{F_\perp(q^2)-\frac{2m_\Lambda (E_N+m_N)}{s_+}(F_\perp(q^2)-F_+(q^2))
-\frac{(E_N-m_\Lambda )(m_\Lambda-m_N)}{q^{2}}(F_0(q^2)-F_+(q^2))\bigg\}\,,\\
\Pi^V_{i 0}(\vec{p}_N)=\,&
\frac{ip^i_N{\cal K}}{2m_N}
\bigg\{F_\perp(q^2)-
\frac{2m_\Lambda (E_N+m_N )}{s_+}(F_\perp(q^2)-F_+(q^2))-\frac{(E_N+m_N )(m_\Lambda-m_N)}{q^{2}}(F_0(q^2)-F_+(q^2))\bigg\}\,,\\
\Pi^A_{j 0}(\vec{p}_N)=\,&\frac{p_N^j{\cal K}}{2m_N}\bigg\{\frac{(m_N+m_\Lambda)(E_N-m_\Lambda)}{q^2}(G_+(q^2)-G_0(q^2))+G_\perp(q^2)-\frac{2m_\Lambda(E_N-m_N)}{s_-}(G_\perp(q^2)-G_+(q^2))\bigg\}\,,\\
\Pi^A_{i j}(\vec{p}_N)=\,&
\frac{-i(m_N+m_\Lambda){\cal K}}{2m_N}
\bigg\{\delta_{ij}\frac{E_N+m_N}{m_N+m_\Lambda}G_\perp(q^2)+\frac{p^i_Np^j_N}{q^2}\left(G_+(q^2)-G_0(q^2)+\frac{2m_\Lambda q^2(G_+(q^2)-G_\perp(q^2))}{s_-(m_N+m_\Lambda)}\right)\bigg\}\,,
\end{align}
}
\end{widetext}
with the kinematic factor defined as
\begin{equation}
 {\cal K}=\sqrt{\frac{2m_N^2}{E_N(E_N+m_N)}}\,. 
\end{equation}
In this transition, the system of equations has rank $3^{\text{V}} + 3^{\text{A}}$, matching the number of independent form factors. Once linearly dependent matrix elements are averaged, the system becomes exactly solvable.
In our analysis, we first perform the decomposition into form factors at each value of $q^2$, and only then apply fits. This order is justified because solving the linear system involves linear combinations of matrix elements, which do not alter the spectral decomposition of the correlation functions, nor the validity of the fitting Ans\"atze used in both the two-state and summation methods.

\begin{table}
    \centering
    \begin{tabular}{ccc}
        \hline
        \hline
        $f_0$ & $f_+$ & $f_\perp$ \\
        \hline
        1.1849(57) & 1.1849(57) & 2.695(36)\\
        $\langle r^2_{F_0} \rangle m_\Lambda^2$ & $\langle r^2_{F_+} \rangle m_\Lambda^2$ & $\langle r^2_{F_\perp} \rangle m_\Lambda^2$ \\
        \hline
        11.45(46) & 11.72(28) & 13.7(13) \\  
        \hline
        \hline
        $g_0$ & $g_+$ & $g_\perp$ \\
        \hline
        0.8178(55) & 0.8178(55) & 0.8308(48) \\
        $\langle r^2_{G_0} \rangle m_\Lambda^2$ & $\langle r^2_{G_+} \rangle m_\Lambda^2$ & $\langle r^2_{G_\perp} \rangle m_\Lambda^2$ \\
        \hline
        26.56(60) & 10.17(77) & 6.96(38)  \\     \hline
        \hline
    \end{tabular}

\begin{align*}
\begin{blockarray}{ccccccc}
~\text{corr.}~~ & f_0 & \langle r^2_{F_0} \rangle & f_+ & \langle r^2_{F_+} \rangle & f_\perp & \langle r^2_{F_\perp} \rangle\\
\begin{block}{c(cccccc)}
f_0&1.00 & -0.36 & 1.00 & 0.07 & 0.15 & -0.11 \\
\langle r^2_{F_0}\rangle&-0.36 & 1.00 & -0.36 & 0.52 & -0.08 & 0.16 \\
f_+&1.00 & -0.36 & 1.00 & 0.07 & 0.15 & -0.11 \\
\langle r^2_{F_+} \rangle&0.07 & 0.52 & 0.07 & 1.00 & 0.03 & 0.14\\
f_\perp&0.15 & -0.08 & 0.15 & 0.03 & 1.00 & 0.71 \\
 \langle r^2_{F_\perp} \rangle&-0.11 & 0.16 & -0.11 & 0.14 & 0.71 & 1.00\\
\end{block}\\
~\text{corr.}~~ & g_0 & \langle r^2_{G_0} \rangle & g_+ & \langle r^2_{G_+} \rangle & g_\perp & \langle r^2_{G_\perp} \rangle\\
\begin{block}{c(cccccc)}
g_0&1.00 & 0.09 & 1.00 & -0.42 & 0.82 & 0.12 \\
\langle r^2_{G_0} \rangle &0.09 & 1.00 & 0.09 & -0.02 & -0.14 & 0.47 \\
g_+&1.00 & 0.09 & 1.00 & -0.42 & 0.82 & 0.12 \\
\langle r^2_{G_+} \rangle&-0.42 & -0.02 & -0.42 & 1.00 & 0.08 & 0.34 \\
g_\perp&0.82 & -0.14 & 0.82 & 0.08 & 1.00 & 0.07 \\
\langle r^2_{G_\perp} \rangle&0.12 & 0.47 & 0.12 & 0.34 & 0.07 & 1.00\\
\end{block}
\end{blockarray}
\end{align*}
    
    \caption{\label{tab:chargeradi_H}Charges and radii of the helicity form factors and the corresponding correlation matrix. The radii are provided in units of the experimental $m_\Lambda^2$ with $m_\Lambda = 1.11568$\,GeV.}
\end{table}

\subsection{The z-expansion}
To parametrize the $q^2$-dependence of the form factors, we employ the model-independent z-expansion~\cite{Hill:2010yb}, defined by
\begin{align}
G(q^2)=\sum_{k=0}^{k_{\max}}a_k z^k~\text{with}~ z=\frac{\sqrt{t_{\rm cut}-q^2}-\sqrt{t_{\rm cut}}}{\sqrt{t_{\rm cut}-q^2}+\sqrt{t_{\rm cut}}},
\end{align}
where $t_{\rm cut}=(m_\pi+m_K)^2$ with $m_\pi$ and $m_K$ denoting the pion and kaon masses, respectively. It is customary~\cite{Hill:2010yb} to constrain the $z$-expansion to vanish in the large-$Q^2$ ($=-q^2$) limit by imposing $a_{k_{\max}} = -\sum_{k=0}^{k_{\max}-1}a_k$. From this parametrization, the charges and radii are extracted as
\begin{equation}
g_i=a_0\quad\text{and}\quad
    \langle r^2_{G_i} \rangle = -\frac{3\alpha_1}{2\alpha_0 t_{\rm cut}}.
\end{equation}

\subsection{Final results}

Our final results for the helicity form factors—subsequently used to derive the Weinberg form factors via conversion formulas, as well as the decay rates and results shown in the letter—are displayed in Fig.~\ref{fig:HFF}. We observe very good agreement between the two-state fit and the summation method; the two are combined with equal weights. For the $q^2$-dependence fits, we use $k_{\max} = 3$ for all form factors and enforce vanishing behavior as $q^2 \to -\infty$ by setting $a_3 = -(a_0 + a_1 + a_2)$. 
Symmetry relations from Eqs.~\eqref{eq:constraint1} and~\eqref{eq:constraint2} are imposed through a combined fit by enforcing
\begin{equation}
\begin{aligned}
f_0 = f_+\,, \quad g_0 = g_+\,, \quad
G_\perp(q^2_{\max}) = G_+(q^2_{\max})\,.  
\end{aligned}    
\end{equation}
The combined fit also allows correlations between form factors to be propagated to the final fit parameters. Our final results, expressed as charges and radii, are listed in Table~\ref{tab:chargeradi_H}, while Table~\ref{tab:zexp_H} lists the parameters of the order-4 $z$-expansion described above with $a_3$ derived implicitely.

\begin{table*}[t]

% ----------- F SET -----------
\begin{minipage}{0.35\textwidth}
\begin{tabular}{ccc}
\hline\hline
        $\alpha^{F_0}_0$ & $\alpha^{F_0}_1$ & $\alpha^{F_0}_2$ \\
        \hline
        1.1849(57) & -2.87(11)  & -1.0(11)\\
        $\alpha^{F_+}_0$ & $\alpha^{F_+}_1$ & $\alpha^{F_+}_2$ \\
        \hline
        1.1849(57) & -2.939(72)  & -1.74(69)\\
        $\alpha^{F_\perp}_0$ & $\alpha^{F_\perp}_1$ & $\alpha^{F_\perp}_2$ \\
        \hline
        2.695(36) & -7.81(81)  & 6.7(61) \\
\hline\hline
\end{tabular}
\end{minipage}
\begin{minipage}{0.60\textwidth}
\centering
\begin{align*}
\begin{blockarray}{cccccccccc}
~\text{corr.}~~ & \alpha^{F_0}_0 & \alpha^{F_0}_1 & \alpha^{F_0}_2 & \alpha^{F_+}_0 & \alpha^{F_+}_1 & \alpha^{F_+}_2 & \alpha^{F_\perp}_0 & \alpha^{F_\perp}_1 & \alpha^{F_\perp}_2 \\
\begin{block}{c(ccccccccc)}
\alpha^{F_0}_0&1.00 & 0.25 & -0.44 & 1.00 & -0.26 & -0.19 & 0.15 & 0.08 & -0.15 \\
\alpha^{F_0}_1 &0.25 & 1.00 & -0.84 & 0.25 & 0.49 & -0.58 & 0.06 & 0.13 & -0.19 \\
\alpha^{F_0}_2 &-0.44 & -0.84 & 1.00 & -0.44 & -0.28 & 0.58 & -0.08 & -0.10 & 0.18 \\
\alpha^{F_+}_0&1.00 & 0.25 & -0.44 & 1.00 & -0.26 & -0.19 & 0.15 & 0.08 & -0.15 \\
\alpha^{F_+}_1&-0.26 & 0.49 & -0.28 & -0.26 & 1.00 & -0.75 & -0.05 & 0.11 & -0.11 \\
\alpha^{F_+}_2&-0.19 & -0.58 & 0.58 & -0.19 & -0.75 & 1.00 & -0.01 & -0.13 & 0.20 \\
\alpha^{F_\perp}_0&0.15 & 0.06 & -0.08 & 0.15 & -0.05 & -0.01 & 1.00 & -0.77 & 0.62 \\
\alpha^{F_\perp}_1&0.08 & 0.13 & -0.10 & 0.08 & 0.11 & -0.13 & -0.77 & 1.00 & -0.97 \\
\alpha^{F_\perp}_2&-0.15 & -0.19 & 0.18 & -0.15 & -0.11 & 0.20 & 0.62 & -0.97 & 1.00 \\
\end{block}
\end{blockarray}
\end{align*}
\end{minipage}

% ----------- G SET -----------
\begin{minipage}{0.35\textwidth}
\begin{tabular}{ccc}
\hline\hline
        $\alpha^{G_0}_0$ & $\alpha^{G_0}_1$ & $\alpha^{G_0}_2$ \\
        \hline
        0.8178(55) & -4.60(11)  & 11.5(10)\\
        $\alpha^{G_+}_0$ & $\alpha^{G_+}_1$ & $\alpha^{G_+}_2$ \\
        \hline
        0.8178(55) & -1.76(13)  & 4.1(13)\\
        $\alpha^{G_\perp}_0$ & $\alpha^{G_\perp}_1$ & $\alpha^{G_\perp}_2$ \\
        \hline
         0.8308(48) & -1.222(68)  & -0.35(65) \\
\hline\hline
\end{tabular}
\end{minipage}
\begin{minipage}{0.60\textwidth}
\centering
\begin{align*}
\begin{blockarray}{cccccccccc}
~\text{corr.}~~ & \alpha^{G_0}_0 & \alpha^{G_0}_1 & \alpha^{G_0}_2 & \alpha^{G_+}_0 & \alpha^{G_+}_1 & \alpha^{G_+}_2 & \alpha^{G_\perp}_0 & \alpha^{G_\perp}_1 & \alpha^{G_\perp}_2 \\
\begin{block}{c(ccccccccc)}
\alpha^{G_0}_0&1.00 & -0.36 & 0.09 & 1.00 & 0.34 & -0.33 & 0.82 & -0.21 & -0.03 \\
\alpha^{G_0}_1 &-0.36 & 1.00 & -0.90 & -0.36 & -0.11 & 0.29 & -0.10 & 0.48 & -0.35 \\
\alpha^{G_0}_2 &0.09 & -0.90 & 1.00 & 0.09 & 0.02 & -0.20 & -0.11 & -0.37 & 0.39 \\
\alpha^{G_+}_0&1.00 & -0.36 & 0.09 & 1.00 & 0.34 & -0.33 & 0.82 & -0.21 & -0.03 \\
\alpha^{G_+}_1&0.34 & -0.11 & 0.02 & 0.34 & 1.00 & -0.75 & -0.16 & 0.37 & -0.27 \\
\alpha^{G_+}_2&-0.33 & 0.29 & -0.20 & -0.33 & -0.75 & 1.00 & 0.07 & -0.27 & 0.22 \\
\alpha^{G_\perp}_0&0.82 & -0.10 & -0.11 & 0.82 & -0.16 & 0.07 & 1.00 & -0.17 & -0.11 \\
\alpha^{G_\perp}_1&-0.21 & 0.48 & -0.37 & -0.21 & 0.37 & -0.27 & -0.17 & 1.00 & -0.84 \\
\alpha^{G_\perp}_2&-0.03 & -0.35 & 0.39 & -0.03 & -0.27 & 0.22 & -0.11 & -0.84 & 1.00 \\
\end{block}
\end{blockarray}
\end{align*}
\end{minipage}

% ----------- F SET -----------
\begin{minipage}{0.35\textwidth}
\centering
\begin{tabular}{ccc}
\hline\hline
$\alpha^{F_1}_0$ & $\alpha^{F_1}_1$ & $\alpha^{F_1}_2$ \\
\hline
1.1849(57) & -2.383(73)  & -2.47(70)\\
$\alpha^{F_2}_0$ & $\alpha^{F_2}_1$ & $\alpha^{F_2}_2$ \\
\hline
 0.821(19) & -2.95(45)  & 5.0(33)\\
$\alpha^{F_3}_0$ & $\alpha^{F_3}_1$ & $\alpha^{F_3}_2$ \\
\hline
0.062(13) & -0.33(14)  & 0.68(45)\\
\hline\hline
\end{tabular}
\end{minipage}
\begin{minipage}{0.60\textwidth}
\centering
\begin{align*}
\begin{blockarray}{cccccccccc}
~\text{corr.}~~ & \alpha^{F_1}_0 & \alpha^{F_1}_1 & \alpha^{F_1}_2 & \alpha^{F_2}_0 & \alpha^{F_2}_1 & \alpha^{F_2}_2 & \alpha^{F_3}_0 & \alpha^{F_3}_1 & \alpha^{F_3}_2 \\
\begin{block}{c(ccccccccc)}
\alpha^{F_1}_0 & 1.00 & -0.26 & -0.20 & -0.03 & 0.11 & -0.13 & -0.50 & 0.44 & -0.39  \\
\alpha^{F_1}_1 & -0.26 & 1.00 & -0.73 & 0.16 & -0.12 & 0.10 & 0.20 & -0.25 & 0.23\\
\alpha^{F_1}_2 &-0.20 & -0.73 & 1.00 & -0.03 & -0.06 & 0.10 & 0.12 & -0.02 & 0.05\\
\alpha^{F_2}_0 & -0.03 & 0.16 & -0.03 & 1.00 & -0.80 & 0.67 & 0.10 & -0.19 & 0.40  \\
\alpha^{F_2}_1 & 0.11 & -0.12 & -0.06 & -0.80 & 1.00 & -0.97 & -0.19 & 0.31 & -0.63 \\
\alpha^{F_2}_2 & -0.13 & 0.10 & 0.10 & 0.67 & -0.97 & 1.00 & 0.21 & -0.34 & 0.68 \\
\alpha^{F_3}_0 & -0.50 & 0.20 & 0.12 & 0.10 & -0.19 & 0.21 & 1.00 & -0.88 & 0.74 \\
\alpha^{F_3}_1 & 0.44 & -0.25 & -0.02 & -0.19 & 0.31 & -0.34 & -0.88 & 1.00 & -0.92 \\
\alpha^{F_3}_2 & -0.39 & 0.23 & 0.05 & 0.40 & -0.63 & 0.68 & 0.74 & -0.92 & 1.00 \\
\end{block}
\end{blockarray}
\end{align*}
\end{minipage}

% ----------- G SET -----------
\begin{minipage}{0.35\textwidth}
\centering
\begin{tabular}{ccc}
\hline\hline
$\alpha^{G_1}_0$ & $\alpha^{G_1}_1$ & $\alpha^{G_1}_2$ \\
\hline
0.8178(55) & -1.119(83)  & -0.49(66) \\
$\alpha^{G_2}_0$ & $\alpha^{G_2}_1$ & $\alpha^{G_2}_2$ \\
\hline
-0.082(19) & 0.66(32)  & -1.0(18) \\
$\alpha^{G_3}_0$ & $\alpha^{G_3}_1$ & $\alpha^{G_3}_2$ \\
\hline
-5.14(11) & 26.9(26)  & -42(20) \\
\hline\hline
\end{tabular}
\end{minipage}
\begin{minipage}{0.60\textwidth}
\centering
\begin{align*}
\begin{blockarray}{cccccccccc}
~\text{corr.}~~ & \alpha^{G_1}_0 & \alpha^{G_1}_1 & \alpha^{G_1}_2 & \alpha^{G_2}_0 & \alpha^{G_2}_1 & \alpha^{G_2}_2 & \alpha^{G_3}_0 & \alpha^{G_3}_1 & \alpha^{G_3}_2 \\
\begin{block}{c(ccccccccc)}
\alpha^{G_1}_0 & 1.00 & -0.34 & 0.03 & 0.50 & -0.27 & 0.13 & -0.17 & 0.02 & 0.02\\
\alpha^{G_1}_1 & -0.34 & 1.00 & -0.85 & -0.50 & 0.57 & -0.45 & 0.02 & 0.27 & -0.32\\
\alpha^{G_1}_2 &0.03 & -0.85 & 1.00 & 0.28 & -0.36 & 0.28 & 0.12 & -0.23 & 0.21\\
\alpha^{G_2}_0 & 0.50 & -0.50 & 0.28 & 1.00 & -0.66 & 0.36 & -0.20 & -0.02 & 0.11\\
\alpha^{G_2}_1 & -0.27 & 0.57 & -0.36 & -0.66 & 1.00 & -0.91 & 0.24 & 0.56 & -0.72 \\
\alpha^{G_2}_2 & 0.13 & -0.45 & 0.28 & 0.36 & -0.91 & 1.00 & -0.19 & -0.78 & 0.93 \\
\alpha^{G_3}_0 &-0.17 & 0.02 & 0.12 & -0.20 & 0.24 & -0.19 & 1.00 & -0.37 & 0.02 \\
\alpha^{G_3}_1 & 0.02 & 0.27 & -0.23 & -0.02 & 0.56 & -0.78 & -0.37 & 1.00 & -0.93\\
\alpha^{G_3}_2 &0.02 & -0.32 & 0.21 & 0.11 & -0.72 & 0.93 & 0.02 & -0.93 & 1.00\\
\end{block}
\end{blockarray}
\end{align*}
\end{minipage}

\caption{
\label{tab:zexp_H}
Coefficients of the $z$-expansion for the axial and vector helicity (top two) and Weinberg (bottom two) form factors, along with their corresponding correlation matrices. The $z$-expansion is of fourth order, with $a_3$ set as $a_3\equiv-(a_0+a_1+a_2)$ so that the form factor vanishes as $q^2 \to -\infty$. We note that the Weinberg form factors can be obtained directly from the helicity form factors by solving Eqs.~\ref{Eq:W3}--\ref{Eq:W8}, but this introduces additional kinematic factors. For convenience, and to obtain a parametrization compatible with a direct $z$-expansion, we resample the resulting curves and fit them by fixing $a_0$ and $a_1$ to exactly reproduce the charges and radii listed in Table~\ref{tab:chargeradi}, while treating $a_2$ as a free parameter and $a_3$ as described above. The resulting curves are fully consistent with the original ones.
}
\end{table*}

\end{document}